\documentclass[journal]{IEEEtran}
\usepackage[T1]{fontenc}
\usepackage{amsmath,cuted}
\usepackage{amsfonts}
\usepackage{multirow}
\usepackage{latexsym}
\usepackage{cite}
\usepackage{color}
\usepackage{pslatex}
\usepackage{xspace}
\usepackage{amssymb}
\usepackage{amsbsy}
\usepackage{eucal}
\usepackage{mathrsfs}
\usepackage{multirow}
\usepackage{setspace}
\usepackage{bm}
\usepackage{cite}
\usepackage{widetext}

\def\1{\ensuremath{_1}}
\def\2{\ensuremath{_2}}
\def\3{\ensuremath{_3}}
\def\4{\ensuremath{_4}}
\def\5{\ensuremath{_5}}
\def\6{\ensuremath{_6}}
\def\7{\ensuremath{_7}}
\def\8{\ensuremath{_8}}

\newcommand{\gae}{\lower 2pt \hbox{$\, \buildrel {\scriptstyle >}\over {\scriptstyle \sim}\,$}}
\newcommand{\lae}{\lower 2pt \hbox{$\, \buildrel {\scriptstyle <}\over {\scriptstyle \sim}\,$}}

\newcommand{\be}{\begin{equation}}
\newcommand{\ee}{\end{equation}}
\newcommand{\benn}{\begin{equation*}}		
\newcommand{\eenn}{\end{equation*}}
\newcommand{\bea}{\begin{eqnarray}}
\newcommand{\eea}{\end{eqnarray}}
\newcommand{\bean}{\begin{eqnarray*}}
\newcommand{\eean}{\end{eqnarray*}}

\newcommand\RR{{\rm I\kern-.16em R}}
\newcommand{\invisible}[1]{ }

\newcommand{\bi}{\begin{itemize}}
\newcommand{\ei}{\end{itemize}}

\newcommand{\Sfr}{\mathfrak{S}}
\newcommand{\xivec}{{\boldsymbol{\xi}}}
\newcommand{\xvec}{{\boldsymbol{x}}}

\usepackage[pdftex]{graphicx}

\begin{document}

\title{Efficient Representation of Uncertainty for Stochastic
  Economic Dispatch}

\author{\IEEEauthorblockN{Cosmin Safta, Richard L.-Y. Chen, Habib N. Najm,
               Ali Pinar}\\
\IEEEauthorblockA{Sandia National Laboratories,  Livermore, CA 94551}\\
\and
\IEEEauthorblockN{Jean-Paul~Watson} \\
\IEEEauthorblockA{Sandia National Laboratories, Albuquerque, NM 87185}\\
Email: \{csafta,rlchen,hnnajm,apinar,jwatson\}@sandia.gov
}

\maketitle

\begin{abstract}
Stochastic economic dispatch models address uncertainties in
forecasts of renewable generation output by considering a finite number
of realizations drawn from a stochastic process model, typically via Monte Carlo
sampling. Accurate evaluations of expectations or higher-order moments for
quantities of interest, e.g., generating cost, can require a prohibitively large
number of samples. We propose an alternative to
Monte Carlo sampling based on Polynomial Chaos expansions. These
representations are based on sparse quadrature methods, and enable accurate
propagation of uncertainties in model parameters. We also investigate a method
based on Karhunen-Loeve expansions that enables us to efficiently represent
uncertainties in renewable energy generation. Considering expected production cost,
we demonstrate that the proposed approach can yield several orders of magnitude
reduction in computational cost for solving stochastic economic dispatch
relative to Monte Carlo sampling, for a given target error threshold.
\end{abstract}

\begin{IEEEkeywords}
Stochastic Economic Dispatch, Monte Carlo Sampling, Polynomial Chaos
Expansion, Karhunen-Loeve Expansion.
\end{IEEEkeywords}
\IEEEpeerreviewmaketitle

\section{Introduction}

\IEEEPARstart{U}{nit} commitment (UC) is the fundamental process of
scheduling thermal generating units in advance of operations in the
electric power grid \cite{Carrion:2006}. The objective is to
minimize overall production costs to satisfy forecasted load for
electricity, while respecting operational constraints of both
transmission elements (e.g., thermal limits) and generators (e.g.,
ramping limits). Economic dispatch (ED) is a closely related
operations problem, in which cost minimization is performed to
identify an optimal set of power output levels for a fixed set of
on-line thermal generating units. Typically, UC and ED are respectively
formulated as mixed-integer and linear optimization problems, and
routinely solved using commercial solvers. Despite recent improvements in
load forecasting technology, next-day predictions are imperfect, with
errors on average in the 1-3\% range and exceeding 10\% on specific days
\cite{isone:2014}. To account for such inaccuracies, reserve margins are
universally imposed in operations planning. These margins implicitly deal
with uncertainty in load forecasts, by ensuring there is sufficient
generation capacity available to meet unexpectedly high load during
operations.

An alternative approach to dealing with forecast errors is to
\emph{explicitly} model the uncertainty, typically via a finite
set of sampled realizations from a stochastic process model. This
approach results in a stochastic ED model (SED), in which the
objective typically is to minimize the expected production cost across
a set of load scenarios \cite{Ruiz:2009,Takriti:96}. By explicitly
representing the inherent uncertainty in load forecasts, a SUC
solution ensures sufficient flexibility to meet a range of potential
realizations during next-day operations. Further, by explicitly
representing uncertainty, reliance on reserve margins is reduced,
yielding less costly solutions than those obtained under
deterministic ED models. Increasing penetration levels of renewables
(e.g., wind and solar) generation accentuate the differences between
stochastic and deterministic grid operation problems, due to increased
errors in next-day forecasts; relative to load, accurate prediction of
next-day meteorological conditions is very difficult.

Despite the conceptual appeal of stochastic variants of grid operations
models, they are not yet used in in practice due to their well-known
computational difficulty \cite{Papavasiliou:2013}. This difficulty is
primarily driven by the number of forecast samples required to achieve
high-quality, robust solutions to models such as stochastic UC and ED.

Uncertainties such as those found in stochastic UC are ubiquitous in
both power systems operations and planning, and the importance of
credibly accounting for them is well-recognized. In their seminal work,
Takriti {\em et al.} introduced a model and solution technique for the
problem of generating electric power when loads are uncertain
\cite{Takriti:96}. More recent work on addressing grid operations under
uncertainty has focused on various
sources of uncertainty and different frameworks for modeling uncertainty
(e.g., stochastic programming, chance constraints, and robust optimization).
For example, \cite{Wang:2013} considers a stochastic UC in which the
availability of consumer demand response (DR) is uncertain. DR uncertainty
was modeled using a set of scenarios, and a chance constraint was imposed
to ensure loss-of-load probability is below a pre-defined risk level.
Chen {\em et al.} \cite{Chen:2014} considered the combined UC and ED
problem under both random and targeted component failures, where
allowable loss-of-load was parameterized by the contingency size. Bertsimas
{\em et al.}\cite{Bertsimas:2013} proposed a two-stage adaptive robust UC
model given uncertainty in nodal net injections, and developed a solution
approach based on a combination of Benders decomposition and
outer approximation.

Despite recent advances, the general lack of advanced methods to accurately
model and represent uncertainty  and the inability of scenario-based approaches
to solve industrial sized stochastic optimization problems have led researchers
to seek alternatives, in both grid operations and planning contexts. For example,
Thiam and DeMarco~\cite{Thiam:2010} argue:
``{\it Simply put, when uncertainty is credibly accounted for such
  methods yield solutions for economic benefit of a transmission
  expansion in which the ``error bars'' are often larger than the
  nominal predicted benefit.}''
In other words, only a limited number of samples could be considered while
sustaining computational tractability, which in turn impacts the ultimate
utility of a solution.
Of course, it is not possible to change the nature of uncertainties, such
that if uncertainties are so large that they fail to provide useful information.
However, as we demonstrate in our computational experiments, it is possible
to significantly reduce the impact of errors introduced by modeling and
sampling of uncertainty.

In this paper, we adopt advanced modeling and sampling techniques from
the uncertainty quantification (UQ) community, and leverage them to impact
power systems operations problems such as stochastic UC and ED. Such
techniques have been successfully applied in many areas of computational
science and engineering, with significant success
(see e.g.,~\cite{Najm:2009a} and references therein). The need for accurate
estimation of uncertain model outputs, along with the prohibitive cost of
requisite large numbers of Monte Carlo (MC) samples, have led to the
development of more efficient alternatives. Specifically, we employ Polynomial
Chaos expansions~\cite{Ghanem:1991} to represent uncertain model inputs
in terms of sets of orthogonal polynomials of standard random variables. The
task of propagating this functional representation to model outputs can be
achieved via several pathways. In our experiments, we employ a Galerkin
projection technique in conjunction with sparse quadrature methods. We
demonstrate that our approach yields a one to two order of  magnitude
reduction in the number of samples (scenarios) required to estimate
expected production cost,relative to MC, depending on given target error
thresholds. Consequently, our approach has the potential to dramatically
reduce the computational difficulty of stochastic grid operations problems,
significantly reducing a major barrier to their use in practice.

The remainder of this paper is organized as follows. We briefly
introduce our stochastic ED formulation in Section~\ref{sec:suc} to
provide context for our research.  In Section~\ref{sec:windkle} we
describe a method to efficiently model the uncertainties in renewables
power production, with an emphasis on wind generation.
Section~\ref{sec:estimation} reviews key concepts in the representation
of uncertainty using Polynomial Chaos and discusses the use of our
surrogate models of renewable resource output for stochastic ED.
We then empirically analyze the accuracy of our surrogates on standard
IEEE test problems in Section~\ref{sec:results}, and conclude in
Section~\ref{sec:conclusion}.

\section{Stochastic Economic Dispatch}
\label{sec:suc}

\IEEEPARstart{W}e abstractly denote the set of unit commitment constraints
(i.e., operational and physical constraints on physical units) as $\mathcal
X$ and let $\xvec \in \mathcal X$ denote a vector of (binary) unit commitment
decisions. In the context of this paper, we assume that the unit
commitment decisions $\boldsymbol x$ are fixed; thus, $\boldsymbol
x$ are parameters in our SED model.

We treat renewable power generation as a random field and present
an efficient approach to represent these random fields in subsequent
sections. In principle, one can also consider uncertain loads and similarly
represent them as random fields  ~\cite{Safta:2014}. For
conciseness, we focus strictly in our experiments on uncertain renewable
outputs and without loss of generality consider deterministic loads.

Because uncertain renewables generation is represented using random fields
and ultimately, as will be illustrated below, as functions of a vector of
random variables $\xivec(\omega)$, the corresponding production cost
$Q(\xvec,\xivec(\omega))$ is similarly uncertain or random. The
\emph{expected} production cost, denoted $\overline Q(\xvec)$, is
defined as
\begin{equation}
\overline Q(\xvec) = \mathrm{E}_{\xivec}
Q(\xvec, \xivec(\omega))
\label{eq:buc}
\end{equation}
\noindent and the (multi-period) \emph{stochastic economic dispatch problem} under a
fixed unit commitment $\xvec$ is given by
\begin{equation}
Q(\xvec, \xivec(\omega))=\min_{\boldsymbol
  {f,p \ge 0,q \ge 0,\theta}} \quad \sum_{t \in T} \sum_{g \in G} c_g^P(p_g^t)+
\sum_{t \in T} \sum_{i \in N} M q_i^t
\label{eq:ed_mod_obj}
\end{equation}
{\noindent \textmd{s.t.}}
\begin{subequations}\label{eq:sed}
\label{eq:rec_obj}
\begin{align}
&\sum_{r \in   R_i}  p_{r}^t(\xivec(\omega))+\sum_{g \in   G_i}  p_g^t
+ \sum_{e \in    E_{.i}}  f_{e}^t - \sum_{e \in    E_{i.}}  f_{e}^t =
D_i^t - q_i^t, \quad \forall i,t\label{eq:ed_pow_bal}\\
&B_e ( \theta_{i}^t -  \theta_{j}^t) -  f_e^t =  0, \quad \forall e = (i,j), t \label{eq:ed_pow_line}\\
&\underline F_e \leq  f_e^t \leq \overline F_e,  \quad  \forall e, t \label{eq:ed_pow_lim}\\
&\underline P_g x_g^t \leq  p_g^t \leq \overline P_g x_g^t, \quad  \forall g, t \label{eq:ed_mod_gen_lim}\\
&p_g^{t} -  p_g^{t-1} \leq R_g^u x_g^{t-1} + {S}_g^u (x_g^t-x_g^{t-1}) + \overline P_g ( 1-x_g^t), \quad  \forall g, t \label{eq:ed_mod_ru}\\
&p_g^{t-1}-  p_g^t\leq R_g^d x_g^{t}+S_g^d (x_g^{t-1}-x_g^t)  +\overline P_g(1-x_g^{t-1}), \quad  \forall g, t \label{eq:ed_mod_rd}
\end{align}
\end{subequations}
\noindent where $R_g^u$ ($R_g^d$) and  $S_g^u$ ($S_g^d$) represent nominal
ramp-up (ramp-down) and  startup (shutdown) rates, respectively.

Here, subscripts $i$ and $j$ denote bus indexes defined over bus set $\mathcal N$, while the superscript $t$ denotes
specific time periods in the planning horizon $t=1,\cdots,T$. Subscript $g$ denotes the generator index defined over generator set $\mathcal G$, while the subscript $e = (i,j)$ denotes the line index (and terminal buses $(i,j)$) defined over transmission line set $\mathcal E$. Renewable generation power (a parameter) is denoted by $p_r^t$ while power from thermal generators (a variable)
is denoted by $p_g^t$. The decision variables $q_i^t$ denotes the load shedding quantity
at bus $i$ at time period $t$. Because renewable generation is a function of RVs, solution
variables are necessarily RVs. The output of interest $Q$ is a function of the same
set of random variables that are used to define the uncertain renewables.

The optimization objective in SED is to minimize the expected total production
and loss-of-load costs. The first term in Eq.~\eqref{eq:ed_mod_obj} represents
total production cost, while the second term represents the loss-of-load penalty.
The load-shedding penalty employs a large positive number $M$, typically
$\$5000$ or $\$6000$ \cite{Papavasiliou:2013}. Eq.~\eqref{eq:sed} specifies
operational and physical constraints on grid components based on a \emph{direct current (DC)} power flow model, and includes (in order):
power balance at each time period and bus~\eqref{eq:ed_pow_bal}; the power flow
$f_e^t$ through each line $e=(i,j)$ as a function of voltage phase angle differences
of the terminal buses, $(\theta_i-\theta_j)$, and the reciprocical of the line reactance,
$B_e$ ~(\ref{eq:ed_pow_line});  lower and upper bounds ($\underline F_e$
and $\overline F_e$, respectively) on line power flow ~\eqref{eq:ed_pow_lim}; lower and upper bounds
($\underline P_g$  and $\overline P_g$, respectively) for the standard
generator power~\eqref{eq:ed_mod_gen_lim}; and generation ramp-up and
ramp-down constraints for pairs of consecutive time
periods~(\ref{eq:ed_mod_ru},
\ref{eq:ed_mod_rd}).

A quadratic production cost function is often employed in economic dispatch
models, e.g., as follows:
\begin{align}
c^P_g(p_g^t) = a_g x_g^t + b_g p_g^t + c_g (p_g^t )^2 \label{quad_prod_cost}.
\end{align}
Equation \eqref{quad_prod_cost} can be accurately approximated using a set
of piecewise linear segments.  For conciseness, we omit these standard
linearization steps. For details concerning linearization of the quadratic
production cost functions, see \cite{Carrion:2006}.

Within the broader context of combined stochastic UC\&ED, the stochastic ED
model is embedded as a sub-problem in the UC model. In a combined stochastic
UC and ED model, the first-stage decisions are the unit commitment selections
$\xvec$, while the optimization objective is to minimize expected production costs.
In the second (recourse) decision stage, uncertain renewable production results in
uncertain (sample-dependent) recourse decisions for the dispatch and load-shedding
variables, and consequently uncertain production and load-shedding costs.
First-stage unit commitment decisions are determined by taking their future impacts
into consideration. These future impacts are quantified by the recourse function
$\overline Q(\xvec)$, which computes the expected value of production cost for a
given (fixed) unit commitment $\xvec$.

Next, we describe a method for representing uncertain wind generation as a
spectral expansion that decouples the deterministic space (time) from the
stochastic space and provides an avenue to significantly reduce the dimensionality
of the stochastic space.

\section{Modeling Wind Uncertainty via Karhunen-Loeve Expansions}
\label{sec:windkle}

Accurate, efficient, and low-dimensional representations of uncertainties
are essential for the success of stochastic grid operations models.
We now discuss how to construct a lower dimensional representation of
wind power uncertainty. Towards this goal, we  explore
representing uncertain wind production time profiles $p_{r}^t$ via Karhunen-Loeve (KL)
expansions~\cite{OlmOmk:2010}. We start in Section~\ref{sec:windhist} by
considering data from NREL's Western Wind Dataset~\cite{nrelwwd:2015},
to assess the feasibility of using KL expansions for representing uncertainty
in wind speed profiles. We then propose, in Section~\ref{sec:windfor},
a model for generating wind power samples that is consistent with
uncertainties observed in current forecast models. We avoid representing
uncertainty in wind power directly, due to the discrete nature of the cut-off
threshold.

\subsection{Karhunen-Loeve Expansions}
\label{sec:windhist}
The KL expansion represents stochastic processes by a linear combination of
orthogonal modes. In order to assess the feasibility of this approach for
representing wind generator output, we consider data for three wind sites
extracted from NREL's Western Wind Dataset~\cite{nrelwwd:2015}.  Two of
these sites are located in Wyoming and are geographically close: site
$\#15414$ located at $(41.48N,105.14W)$ and site $\#16238$ located at
$(41.61N,105.11W)$. The third site, $\#3560$ located at $(35.08N,118.41W)$
in California, was selected to be far from the first two sites. The proximity of
the first two sites allows us to explore correlations in wind power, while the third
site is sufficiently distant from the first two sites such that no spatial correlations
in their production exist.

\begin{figure}[t!]
  \centering
  \includegraphics[width=0.4\textwidth]{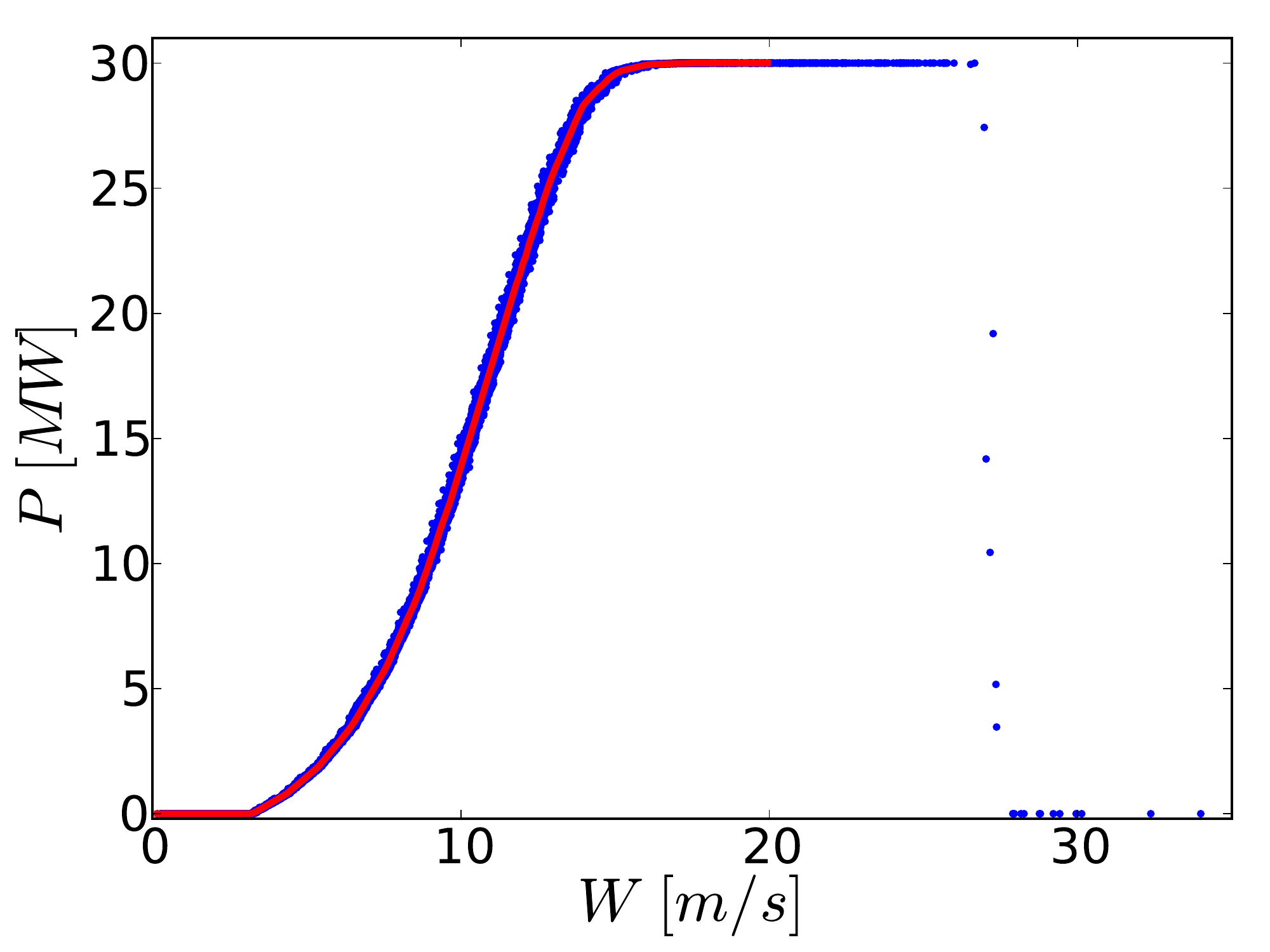}
  \caption{\label{fig:wdata} Rated power output $P$ vs. wind speed $W$ at
    site $\#15414$ for year 2004. Blue dots represent instantaneous
    measurements while the red curve represents power data
    filtered with a top-hat filter of width $\Delta_W=0.05$ [m/s], and
    interpolated via cubic splines.}
\end{figure}

For each site, wind speed and power data is available at 10 minute
intervals for the years 2004 through 2006. Fig.~\ref{fig:wdata} shows
a scatter plot of rated power output vs. wind speed at wind site \#15414.
The rated power output is zero for wind speeds smaller than a threshold
value of approximately  3.2 m/s. At speeds greater then approximately
26 m/s the generators are turned off, for safety reasons.

\begin{figure}[t!]
  \centering
  \includegraphics[width=0.4\textwidth]{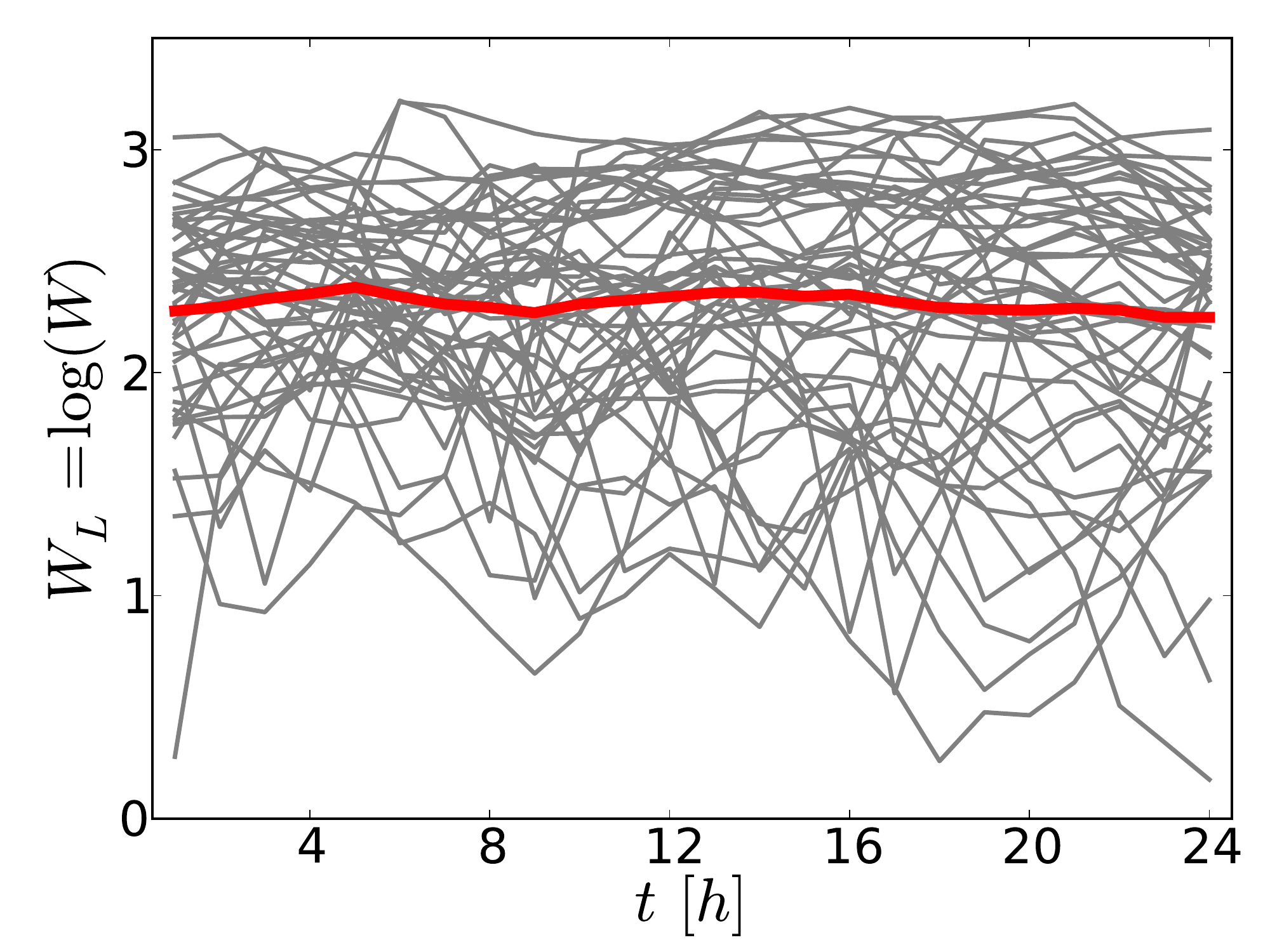}
  \caption{Select daily wind data from January 2004-2006 at
    site $\#15414$. The red curve corresponds to the mean of these daily
    samples.\label{fig:wspl}}
\end{figure}

We take the following steps to post-process the NREL wind data into
wind samples for use in our SED model:
\begin{enumerate}
\item Construct hourly averages for the wind speed; consider wind
  speed data for each day to be an independent sample from a
  24-dimensional random field.
\item Choose a range of dates and assemble a set of 24-dimensional
  samples from this date range. For example, considering the month of
  January for 2004 through 2006 leads to 93 samples. This step was
  adopted to account for seasonal changes in wind patterns.
\end{enumerate}
Fig.~\ref{fig:wspl} shows select daily log-transformed ($W_L=\log(W)$)
wind speed samples at site $\#15414$. We adopted this transformation to
ensure positivity of wind speed samples, generated via the algorithm
described below.

We represent $W_L$ via a KL expansion
\begin{equation}
W_L(t,\omega) = \left < W_L(t,\omega) \right >
            + \sum_{k=1}^{\infty} \sqrt{\lambda_k} f_k(t) \xi_k(\omega)
\label{eq:kle}
\end{equation}
where  $\left< W_L(t,\omega) \right >$  denotes the mean of
$W_L(t,\omega)$,
$f_k(t)$ and $\lambda_k$ are respectively eigenfunctions (eigenvectors
in a discrete setting) and eigenvalues of the covariance matrix
$\Sigma_{W_L}$ of $W_L(t,\omega)$, and $\xi_k$ denotes uncorrelated
random variables with zero mean and unit variance. Projecting
realizations of $W_L$ onto $f_k$ leads to samples of $\xi_k$.
These samples are generally not independent. In the special case where
$W_L$ is a Gaussian random process, $\xi_k$ are i.i.d. standard normal
random variables.

If known, the covariance matrix $\Sigma_W$ can be specified
analytically. Otherwise, if sufficient samples are available, the
covariance matrix can also be estimated from these realizations. For
In our experiments, we estimate the covariance matrix using daily samples
for select times of the year during 2004-2006. Once the covariance matrix
is available, the eigenvalues and eigenvectors in Eq.~(\ref{eq:kle})
are given by solutions of the Fredholm equation of second kind
\[
\int\Sigma_{W_L}(t,s) f(s)\,ds=\lambda f(t).
\]
We discretize and solve this equation using the Nystr\"{o}m
method~\cite{Nystrom:1930}, considering a mid-point quadrature rule
to integrate over the discrete 24-hour period.

\begin{figure}[t!]
  \centering
    \includegraphics[width=0.24\textwidth]{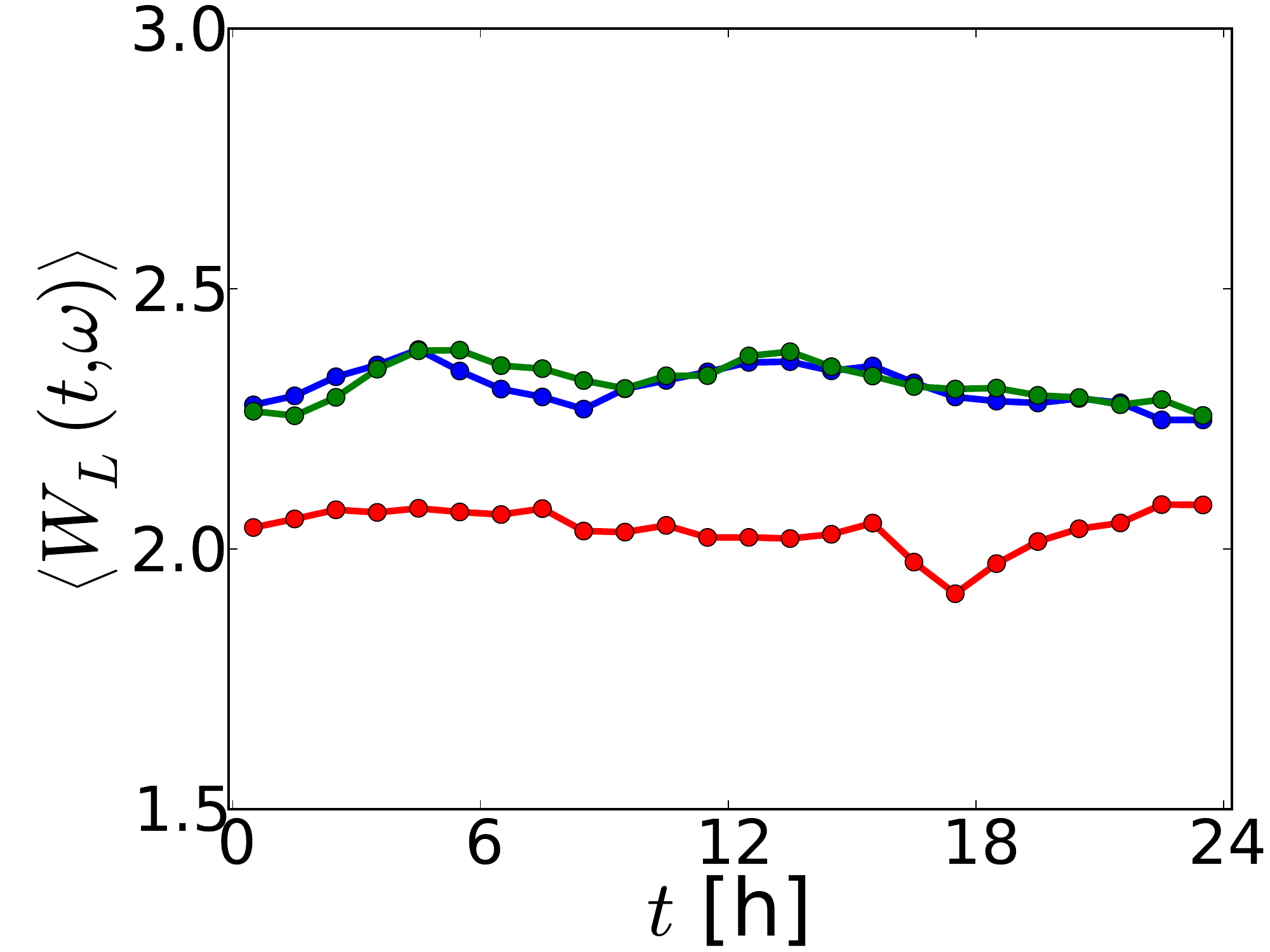}
    \includegraphics[width=0.24\textwidth]{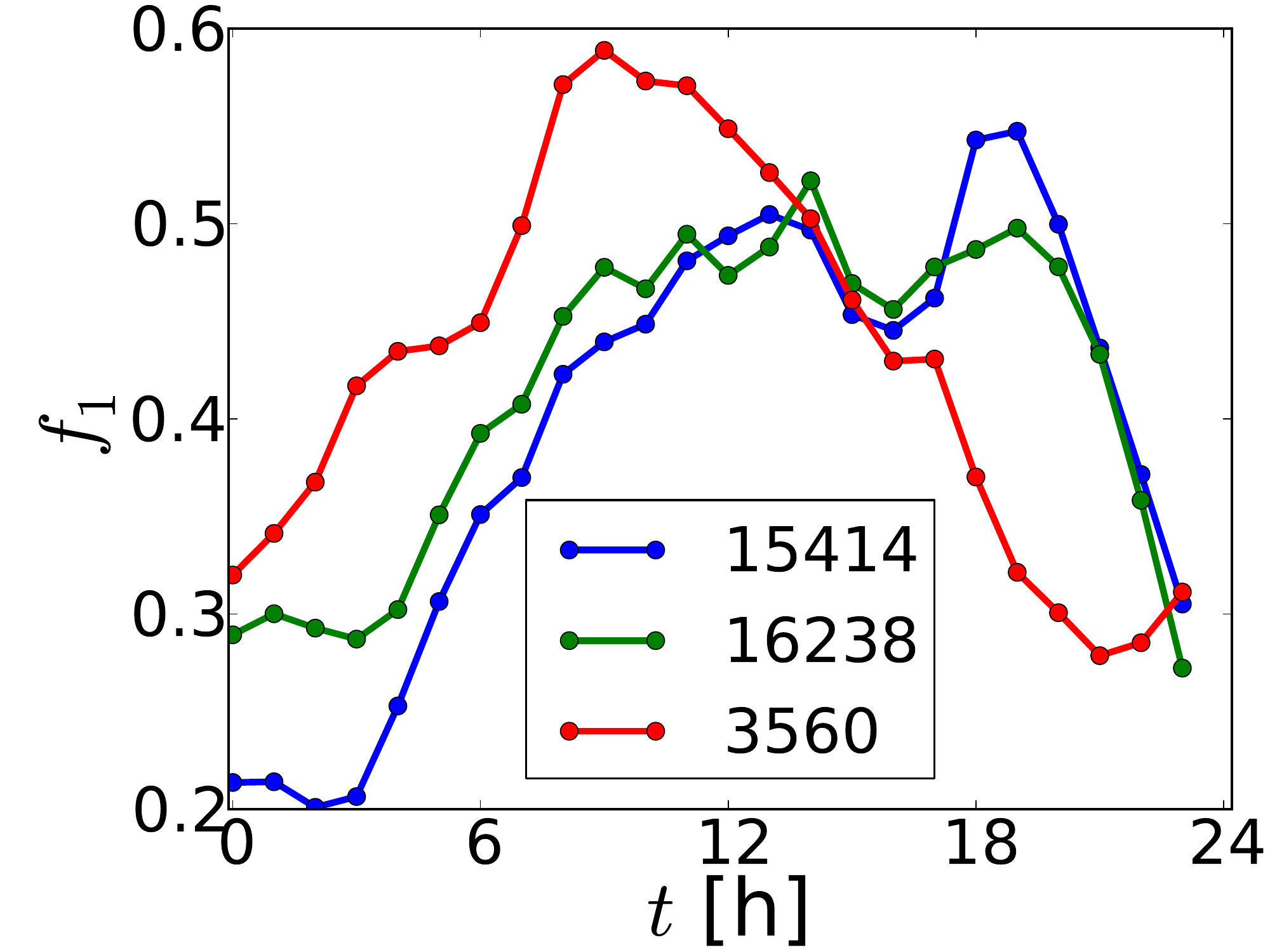}\\
    \includegraphics[width=0.24\textwidth]{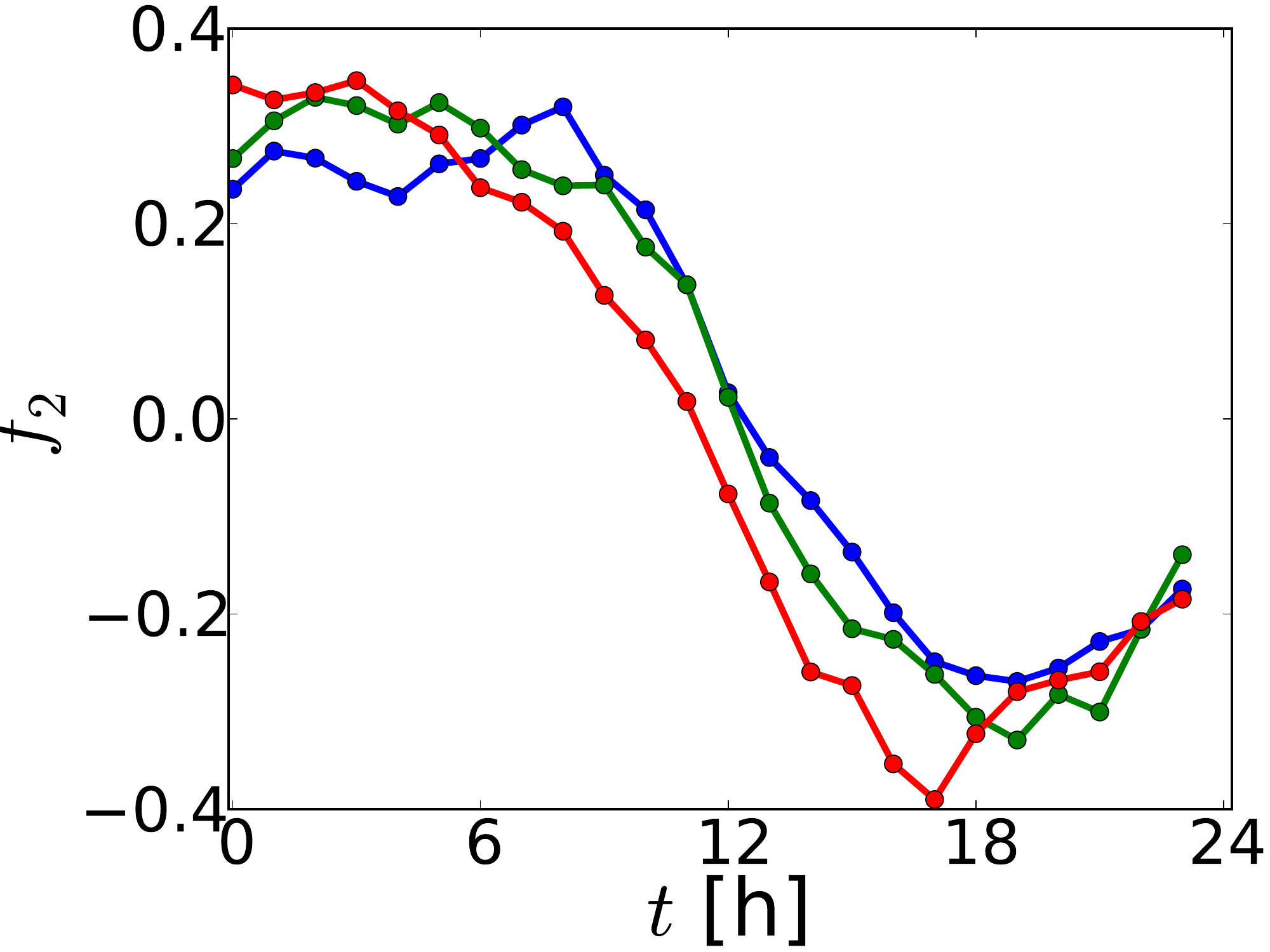}
    \includegraphics[width=0.24\textwidth]{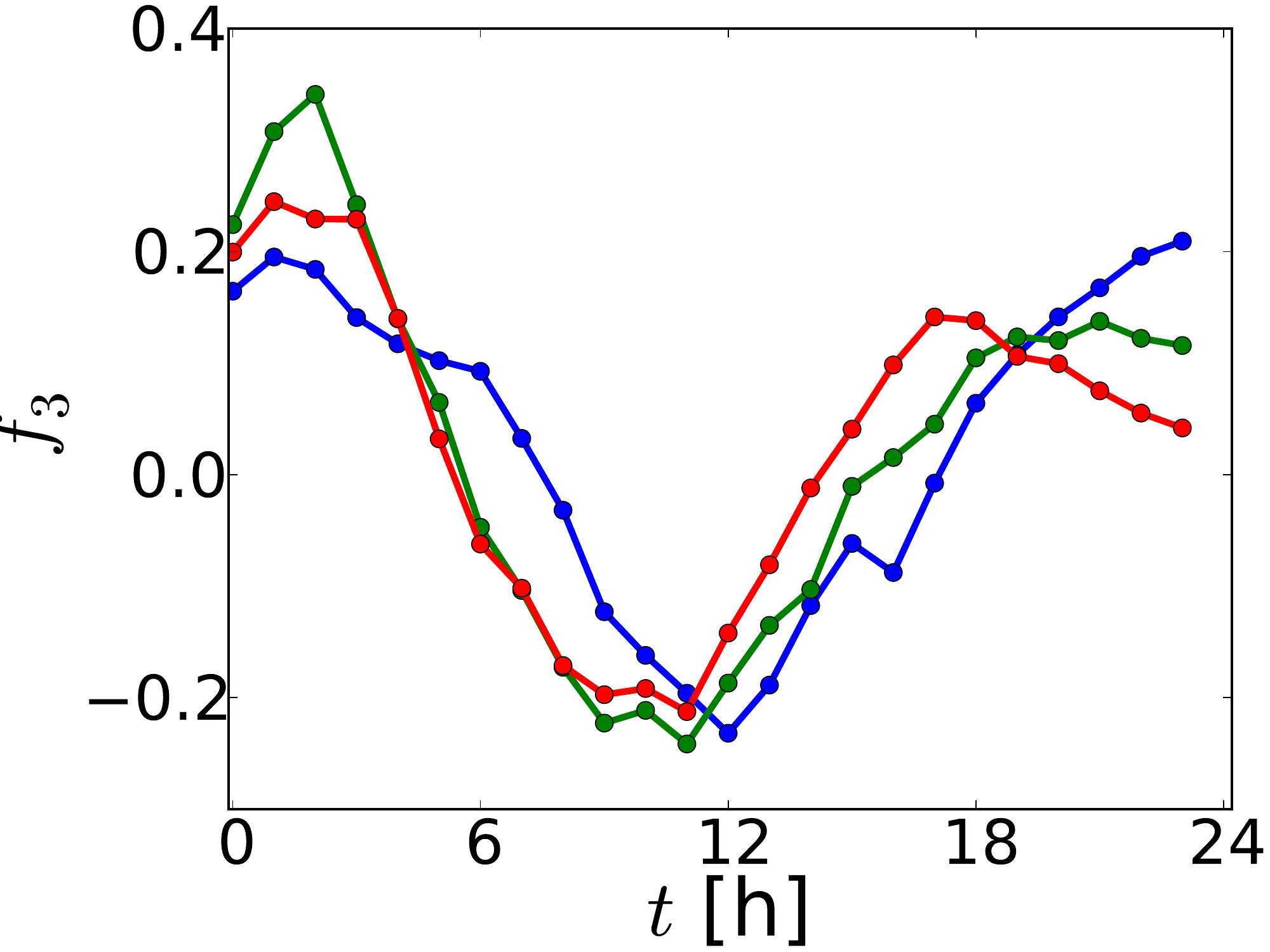}
  \caption{The mean, $\left < W_L(t,\omega) \right >$, and
    first three eigenvectors, $f_1$, $f_2$, and $f_3$,  at several
    wind sites based on data from January 2004-2006.\label{fig:KLmodes}}
\end{figure}

Fig.~\ref{fig:KLmodes} shows the mean and first few KL modes
corresponding to the month of January at the three wind sites
considered in this study. The large degree of similarity for the KL
modes between the three sites suggest a similar structure for the time
correlations in the wind speed. We will further explore the structure of the
covariance matrices $\Sigma_W$ in Section~\ref{sec:windfor}.

\begin{figure}[t!]
  \centering
    \includegraphics[width=0.4\textwidth]{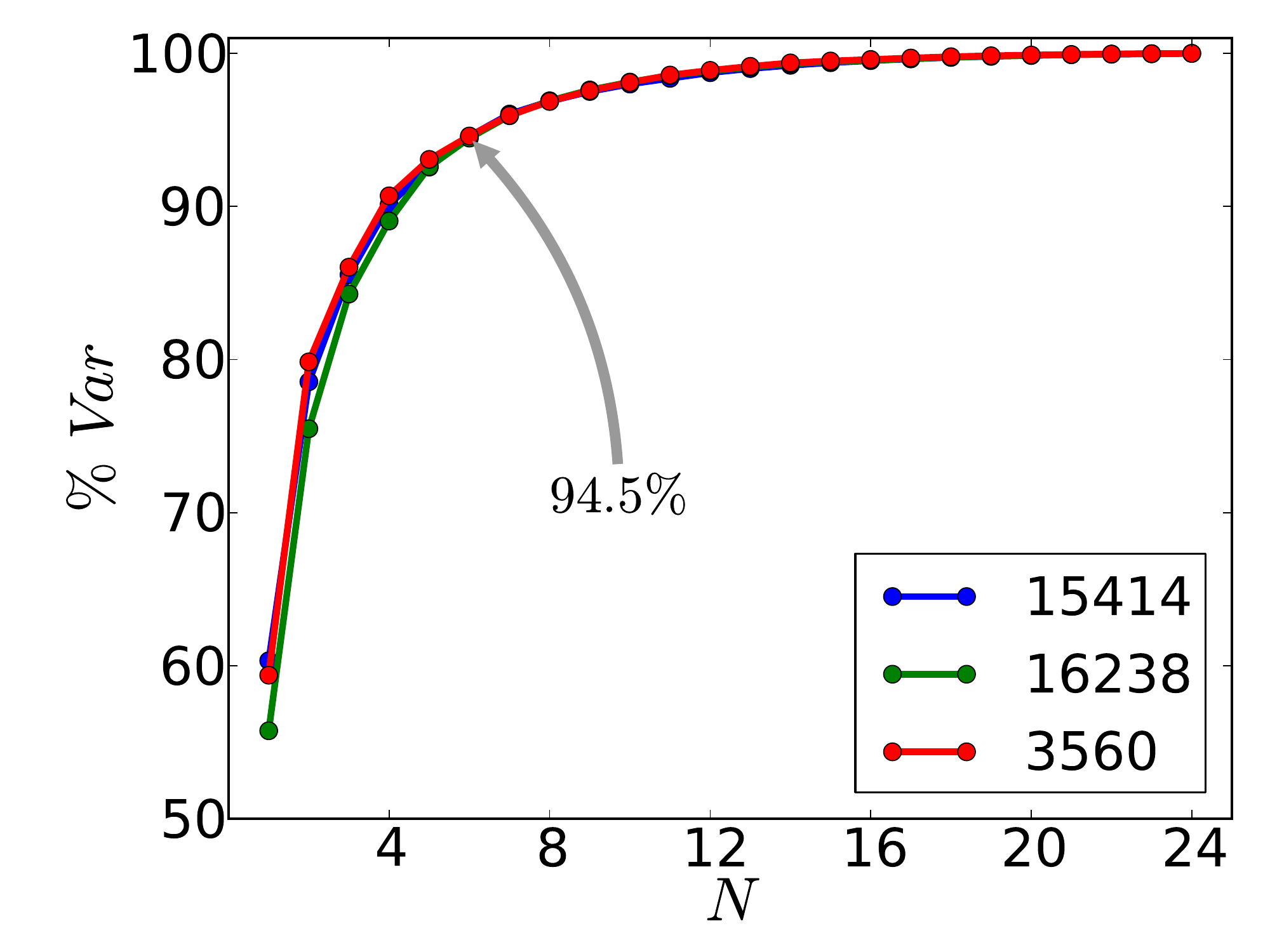}
  \caption{Percentages of the total variance explained by truncated KL expansions
    with an increasing number of modes, for our three wind sites.\label{fig:eig}}
\end{figure}

\begin{figure}[t!]
  \centering
    \includegraphics[width=0.24\textwidth]{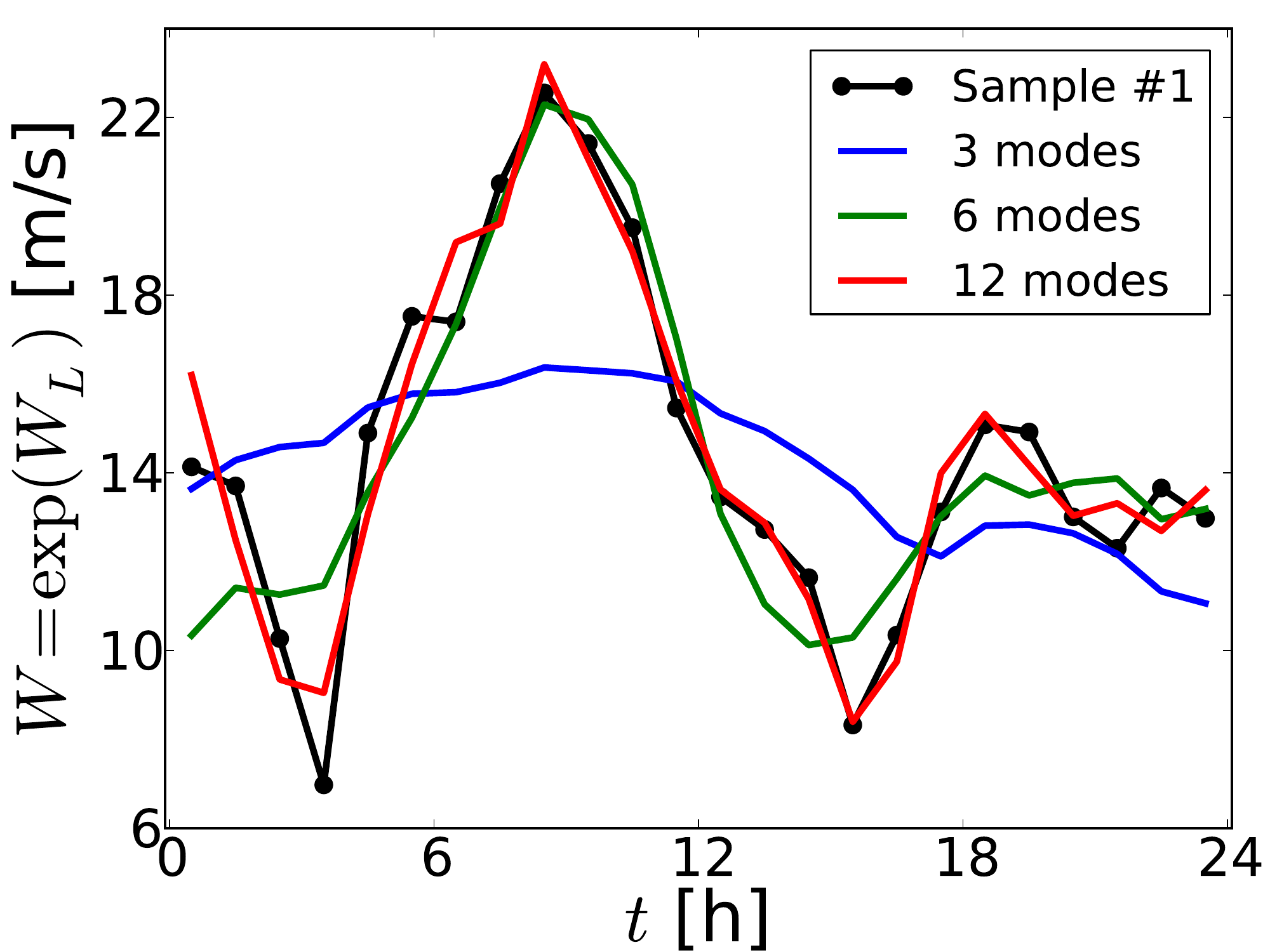}
    \includegraphics[width=0.24\textwidth]{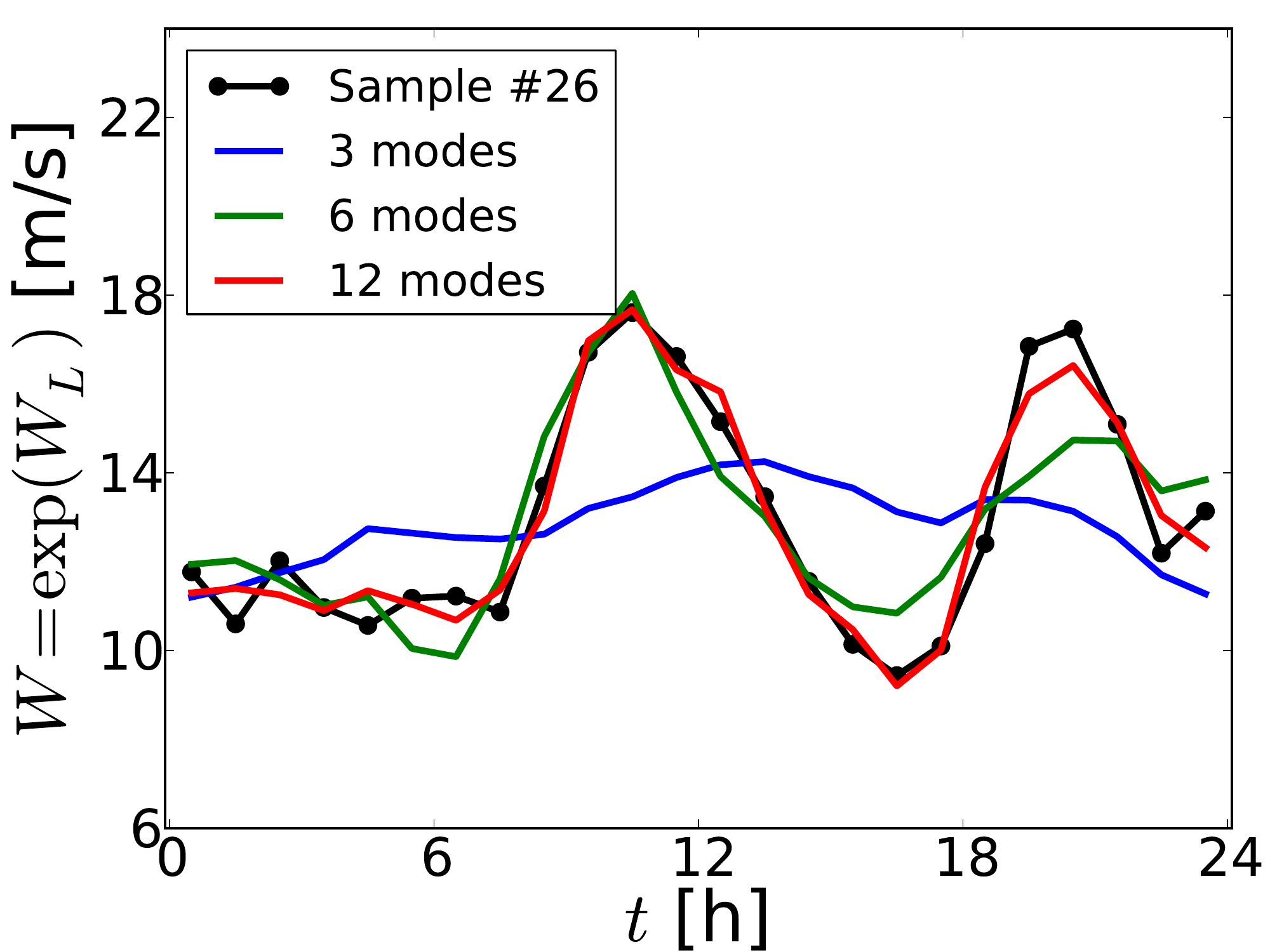}
  \caption{Reconstruction of daily wind samples via truncated
    KL expansion. Left plot corresponds to Jan 1, 2004 and the right plot to
    Jan 26, 2004.\label{eq:KLrec}}
\end{figure}

Because the stochastic dimensions are uncorrelated, the total variance
of the stochastic process is given as the sum of variances from individual
KL mode, as follows:
\[
\textrm{Var}\left[W_L(t,\omega)\right]=\sum_{k=1}^{24}\lambda_k
f_k^2(t)\underbrace{\textrm{Var}[\xi_k]}_{=1}.
\]
Given that eigenvectors are orthonormal with respect to the deterministic
space (the discretized time axis in our study), it follows that:
\be
\int_T \textrm{Var}[W_L(t,\omega)]dt=\sum_{k=1}^{24}\lambda_k.
\ee
As a result, an N-truncated expansion, $N\leq 24$, will explain
\be
100\times\left(\sum_{k=1}^N\lambda_k\right) \Bigg/
\left(\sum_{k=1}^{24}\lambda_k\right)\,[\%]
\label{eq:varfrac}
\ee
of the total variance of the random field. Fig.~\ref{fig:eig} shows
the dependence of the fractional variance given in
Eq.~\eqref{eq:varfrac} on the number of terms $N$ in the truncated KL
expansions at the three wind sites. It is evident that for all sites, $N=6$
modes are sufficient to capture approximately $95\%$ of the total
variance in the daily wind profiles.
We also examine the re-construction of select daily wind samples with
truncated KL expansions, shown in Fig.~\ref{eq:KLrec}. The results indicate
that $N=6$ KL modes are sufficient to represent most of the daily variability.
Similar results are observed for other, randomly selected, samples.

\begin{figure}[t!]
  \centering
    \includegraphics[width=0.24\textwidth]{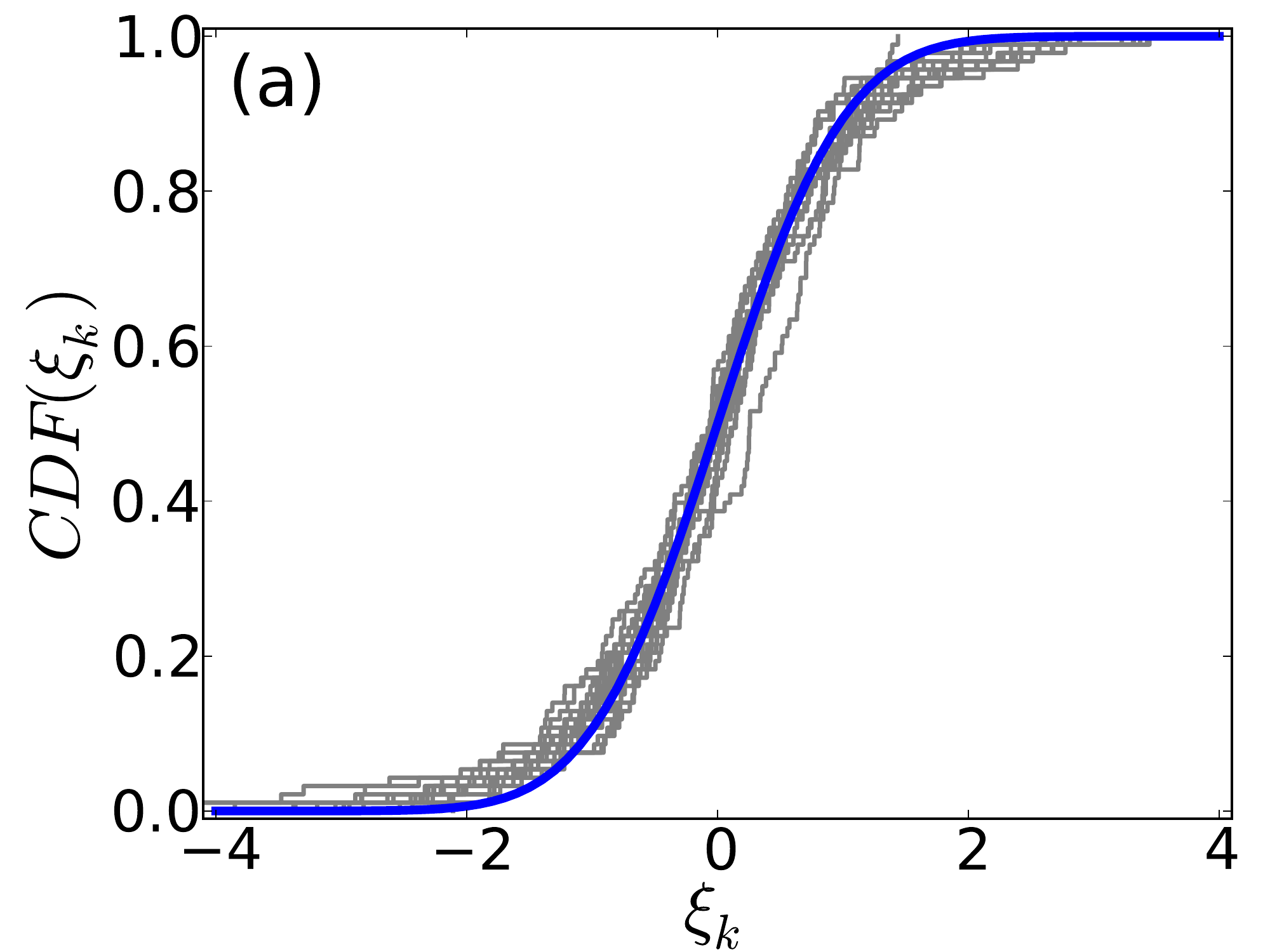}
    \includegraphics[width=0.24\textwidth]{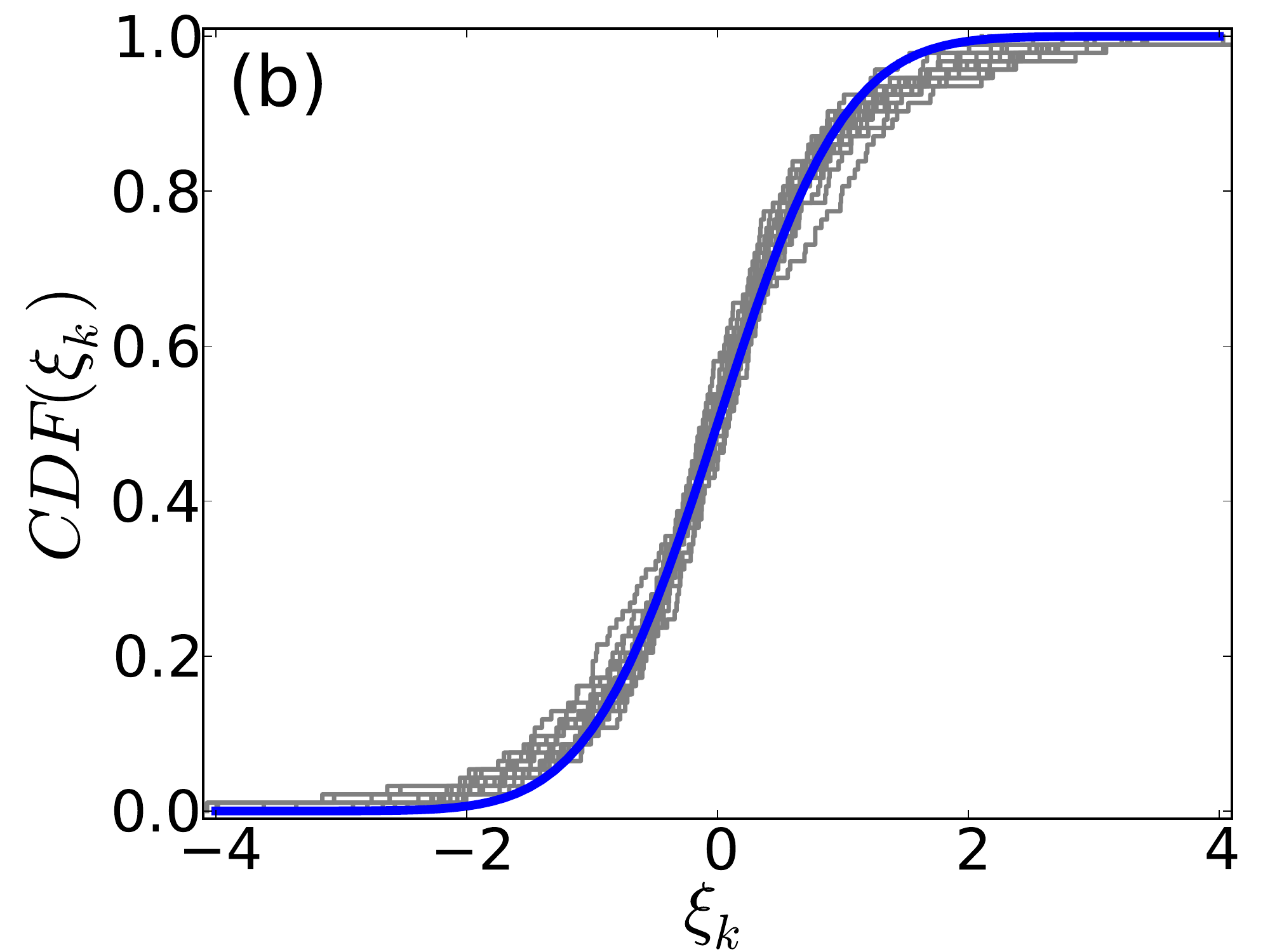}
  \caption{Empirical CDFs (depicted in gray) for $\xi_1$ through $\xi_{15}$
    at (a) site 16238 and (b) site 3560. The CDF for a standard normal random
    variable is shown in blue, for reference.
    \label{fig:rvcdfs}}
\end{figure}

Next, we test the degree of dependence between KL random variables. We
start by constructing empirical cumulative distribution functions (CDFs) from
RV samples. These samples are obtained by projecting each daily wind speed
sample onto the corresponding KL mode. Fig.~\ref{fig:rvcdfs} shows the
empirical CDFs for $\xi_1$ through $\xi_{15}$ at two select wind sites.
Visual inspection of these results and comparison with the CDF of a standard
normal RV indicate strong similarities. Because these RVs are uncorrelated
by construction, modeling them as standard normals implies that they are also
independent.

We further investigate relationships among the standard normal RVs
across the sites we consider in our experiments. To this end we employ
distance correlation factors~\cite{Szekely:2007,Safta:2014b} between
pairs of RVs corresponding to the same KL mode at different sites.
These factors are shown in Table~\ref{tab:dcxi}. Smaller
values, towards zero, indicate negligible dependence between pairs of
RVs, while larger values, close to 1, indicate a strong
dependence. These results indicate a strong correlation between sites
$\#15414$ and $\#16238$ for the first two KL modes, while the same KL
mode at site $\#3560$ shows little correlation with the first two sites. The
third and fourth RVs, as well as the other RVs (not shown), show little
correlation between sites. This is somewhat to be expected given the
nature of turbulence. Specifically, low KL modes are associated with
large scale structures which are likely similar if sites are geographically
situated, while higher order KL modes are associated with smaller
scale structures with much faster eddy turnover times.
\begin{table}[t]
\caption{Distance correlation factors for $\xi_1$ - $\xi_4$
  between wind sites.\label{tab:dcxi}}
\centering
\begin{tabular}{|r|c|c|c|} \hline
RV & 16238-15414 & 16238-3560 & 15414-3560 \\ \hline
$\xi_1$ & 0.91 & 0.24 & 0.24 \\
$\xi_2$ & 0.84 & 0.13 & 0.18 \\
$\xi_3$ & 0.67 & 0.18 & 0.19 \\
$\xi_4$ & 0.53 & 0.23 & 0.22 \\ \hline
\end{tabular}
\end{table}

In experiments described below, we consider a stochastic space with
$N=6$ dimensions at each site. Given that we model the first two modes
at two sites as dependent, the total dimensionality of the stochastic space
is $3\times 6-2=16$. Given the representation of daily wind profiles as
truncated KL expansions, it follows that renewable generation
$p_r^t(\xivec)=f(\exp(W_L(t,\xivec))$ is a function on the same
stochastic space. We neglect the (small) noise in the conversion of wind
speed into power and approximate $f$ as a cubic spline interpolation through
the filtered rated power data. This approximation is depicted by the red
line in Fig.~\ref{fig:wdata}.

\subsection{Wind Power Forecast Models}
\label{sec:windfor}

In this section we discuss the algorithm for generating wind power scenarios
for day-ahead economic dispatch studies. We employ the KLE approach
described in the previous section, with the covariance matrix adjusted to account
for typical uncertainties of wind speed data obtained from weather forecast moedls.

\begin{figure}[t!]
  \centering
    \includegraphics[width=0.45\textwidth]{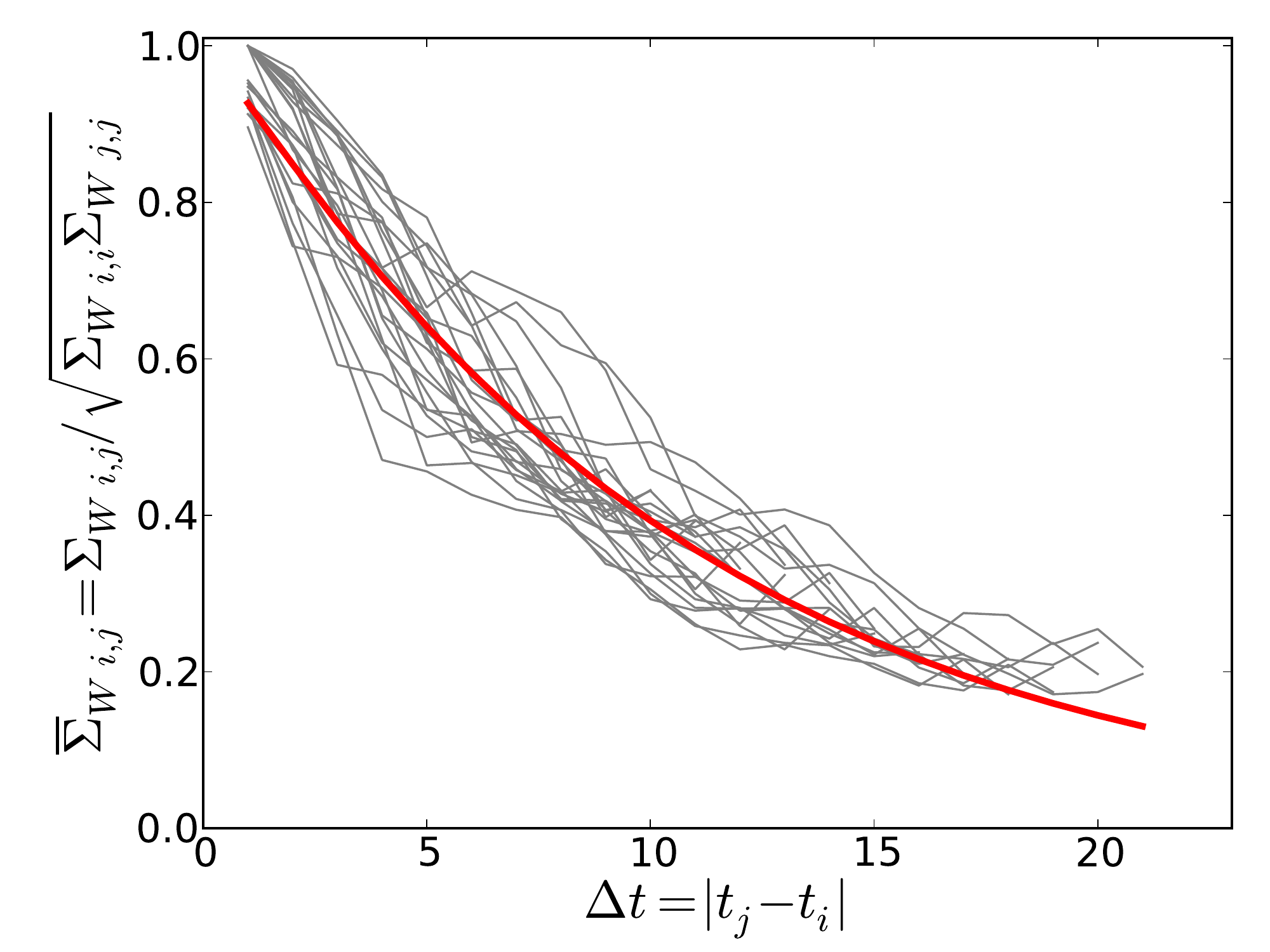}
   \caption{Decay of covariance matrix $\Sigma_W$ components, shown with thin
   grey lines, with increasing time lag $\Delta t=\vert t_j-t_i\vert$,
   anchored at several time instances $i$ for site \#16238. Values are
   normalized by the diagonal entries. Matern
   covariance model is shown with blue line.\label{fig:mfit}}
\end{figure}

We first start by fitting a functional form through the covariance matrix
corresponding to each site. Fig.~\ref{fig:mfit} shows, with thin grey lines,
a typical decay in the components of $\Sigma_W$ with increasing time lag.
We model this data with a Matern covariance kernel~\cite{Rasmussen:2006}
\begin{equation}
  \overline{\Sigma}_W(\Delta t)=\frac{2^{1-\nu}}{\Gamma(\nu)}
  \left(\frac{\sqrt{2\nu}\Delta t}{l_t}\right)K_\nu
  \left(\frac{\sqrt{2\nu}\Delta t}{l_t}\right).
  \label{eq:mcov}
\end{equation}
Here, $\nu$ and $l_t$ are positive parameters, $\Gamma(\cdot)$ is the gamma
function, and $K_\nu(\cdot)$ is the modified
Bessel function of the second kind. The Matern kernel offers more flexibility
for modeling covariances compared to, for example, exponential and square
exponential forms which are particular cases of the Matern kernel, for
$\nu=1/2$ and $\nu\rightarrow\infty$, respectively.
Table~\ref{tab:mcov} shows the parameters of the Matern kernels for the three
sites employed in this study. We obtain similar values for the two sites
(15414 and 16238) that are geographically close. Furthermore, the values for
the $\nu$ parameter for these two sites are close to $\nu=1/2$ indicating
covariance matrices that are close to exponential form.
\begin{table}[t]
\caption{Parameters of Matern kernels corresponding to select
wind sites.\label{tab:mcov}}
\centering
\begin{tabular}{|l|c|c|} \hline
Wind Site & $l_t$ & $\nu$\\ \hline
$15414$ & 11.40 & 0.56 \\
$16238$ & 11.15 & 0.57 \\
$3560$ & 9.79 & 0.78 \\\hline
\end{tabular}
\end{table}
Next we estimate the magnitude of the variance in the estimates of wind
magnitude. Since data on wind forecast and actual realizations
were not immediately available to us, we proceeded to estimate uncertainty
in the wind forecasts given available data for wind power. Specifically,
we used data for day-ahead forecast and actual realizations obtained from
Belgium Electricity Grid Operator ELIA~\cite{eliawwd:2015}. We considered
the aggregate wind power values for the land-based wind farms and computed the
standard deviation for hourly wind power output based on data for years 2012
through 2015. We found a relative value of about $\sigma_P=35\%$. Further, this
value is independent of the time of the day for the day-ahead forecasts.

Based on the formulations presented in this section we employ the following
algorithm to generate wind power samples that are consistent to historical
wind characteristics at the selected wind sites:
\begin{enumerate}
  \item Forecast day-ahead wind profiles at selected wind sites.
  \item Using $\sigma_P$ defined above, estimate a mean $\sigma_W$
  based on the mean forecast wind speed and the rated power output
  curves for each site, similar to the one presented in Fig.~\ref{fig:wdata}.
  \item Construct covariance matrix, ${\Sigma}_W=\sigma_W^2\overline{\Sigma}_W$,
  where $\overline{\Sigma}_W$ is the normalized Matern covariance expression
  presented in Eq.~\eqref{eq:mcov}.
  \item Perform eigen-decomposition for $\Sigma_W$ and generate
  wind samples via Eq.~\eqref{eq:kle} for select samples of $\boldsymbol{\xi}$.
  \item Convert wind samples into wind power values, see Fig.~\ref{fig:wdata}.
\end{enumerate}

\section{Accurate estimation of expected cost with limited samples}
\label{sec:estimation}

We now return to the evaluation of the expected cost in
Eq.~\eqref{eq:buc} corresponding to the ED problem in
Eqs.~\eqref{eq:ed_mod_obj} and \eqref{eq:sed}. We can \emph{estimate}
the expected production cost by using a finite number of
renewable power realizations (i.e., scenarios) $s \in \mathcal S$ sampled from the
joint density $PDF(\xivec)$. For our current example, each $\xi$ is a
standard normal RV, hence the sampling can be done independently in
each stochastic direction. Defining $\rho \equiv 1/|\mathcal{S}|$,
where $|\mathcal{S}|$ is the cardinality of $\mathcal{S}$,
the stochastic ED in Eq.~\eqref{eq:buc} can be rewritten as:
\begin{align}
\min_{\boldsymbol {f, p, q, \theta}} \quad  & \rho \sum_{s \in
  \mathcal S} Q(\boldsymbol x,s) \label{eq:2_stage2}
\end{align}
where $Q(\boldsymbol x,s)$ is the solution of
Eqs.~\eqref{eq:ed_mod_obj} and \eqref{eq:sed} for a particular
instance of the renewable generation $p_r^t(\xivec)\rightarrow p_r^t(s)$.

Formulation \eqref{eq:2_stage2} represents an \emph{extensive form} of
the stochastic ED problem, based on $|\mathcal{S}|$ sampled scenarios
from the stochastic space corresponding to wind power generators.
The typical scenario sampling approach described above uses
Monte Carlo (MC) sampling to approximate an integration, thereby estimating an
expectation.  While  MC algorithms are commonly used for their convenience and
robustness, their poor convergence rate is well-known.  The MC estimate of the
expectation has error
\begin{equation}
\mathrm{V}[Q(\xvec, \xivec)]/\sqrt{|S|},
\label{eq:error}
\end{equation}
where $\mathrm{V[Q]}$ denotes the variance of the RV $Q$.  Given the significant
additional complexity incurred by including stochasticity in the optimization
problem, a stochastic  formulation becomes advantageous relative to a deterministic
formulation when the variance is large.  Hence, accurate estimation of the
expectation is not only an academic exercise but is important in practice.

According to Eq.~\eqref{eq:error}, accurate estimation can be achieved
by increasing the number of samples. However, a linear decrease in
error requires a quadratic increase in the number of samples, which
can quickly render the stochastic optimization problem
intractable. This illustrates the limitation of MC algorithms in
providing accurate estimations; while they are convenient, they are
not efficient.

We propose a method based on Polynomial Chaos expansions that can
enable  high precision quantification of uncertainties with fewer
samples.  In the following sections, we will first  outline the Polynomial Chaos
expansion construction and then present its implementation for the ED problem.

\subsection{Representation of uncertainty using Polynomial Chaos}

Given the formulation in Eq.~(\ref{eq:rec_obj}) with uncertain/random loads
leading to uncertain/random production costs, we employ efficient UQ methods
that rely on \emph{functional representations} of random variables.  Specifically,
we use Polynomial Chaos (PC) expansions. A brief description of PC is presented
below. For an in-depth description, the reader is referred to a series of
publications on this topic~\cite{Wiener:1938,Ghanem:1991,Janson:1997,Xiu:2002c}.

We begin by setting up a requisite theoretical framework as follows.
Define the probability space $(\Omega, \Sfr, P)$, where $\Omega$ is a sample
space, $\Sfr$ is a $\sigma$-algebra on $\Omega$, and $P$ is a probability
measure on $(\Omega,\Sfr)$. Further, defining the \emph{germ}
$\xivec=\{\xi_1,\xi_2,\ldots,\xi_n\}$ as a set of independent
identically distributed (\emph{iid}) RVs in $L_2(\Omega,\Sfr,P)$, to
be further specified below, we focus on the probability space
$(\Omega,\Sfr_\xivec,P)$ employing the sigma algebra generated by $\xivec$.
In this framework, any RV $X: \Omega \rightarrow
\mathbb{R}$, where by construction $X\in L_2(\Omega,\Sfr_\xivec,P)$, can be
written as a PC expansion (PCE):
\be
X(\omega) = X(\xivec(\omega)) = \sum_{k=0}^{\infty} \alpha_k \Psi_k(\xivec)
\label{eq:pcedef}
\ee
where the basis functions $\Psi_k$ are multivariate
polynomials\footnote{Generally, other, non-polynomial basis functions can be
used, but here we restrict ourselves, without loss of generality, to the most
common polynomial-based usage.} that are orthogonal, by construction,
with respect to the density of $\xivec$. Thus
\be
\langle \Psi_i\Psi_j\rangle =
\int\Psi_i(\xi)\Psi_j(\xi)dP(\xi)=\delta_{ij}\langle \Psi_i^2\rangle
\ee
where $\delta_{ij}$ is Kronecker's delta.
Further, given this orthogonality, we have
\be
\alpha_k = \frac{\langle X\Psi_k \rangle}{\langle\Psi_k^2\rangle}
\ee
where the inner product is defined, for any RV $Z(\xivec)$, by the Galerkin
projection
\be
\langle Z \rangle = \int Z(\xivec) p_\xivec(\xivec) d\xivec .
\label{eq:proj}
\ee
Moreover, the $\Psi_k$ are products of univariate polynomials, namely
$\Psi_k(\xivec) =\psi_{k_1}(\xi_1)\cdots \psi_{k_n}(\xi_n)$. In a
practical computational context, one truncates the PCE to order $p$.
The number of terms in the resulting finite PCE
\be
X \approx \sum_{k=0}^{P} \alpha_k \Psi_k(\xivec)
\ee
is given by $P+1={(n+p)!}/{n!p!}$. We dispense with the $\approx$ symbol in the
remainder of this paper, employing for any RV $X(\xivec)$ its truncated PCE
\be
X = \sum_{k=0}^{P} \alpha_k \Psi_k(\xivec).
\label{eq:pce}
\ee
Generalized PC (gPC) expansions have been developed by Xiu and
Karniadakis~\cite{Xiu:2002c} using a
broad class of orthogonal polynomials in the ``Askey'' scheme~\cite{Askey:1985}.
Each family of polynomials corresponds to a given choice of distribution for the
$\xi_i$ and is, by construction, orthogonal with respect to the density of $\xi_i$.
In general, the most useful choices for $(\xivec,\Psi)$ are normal RVs
with Hermite polynomials and uniform RVs with Legendre polynomials.

\subsection{Construction of PCE for the Minimum Cost}

In the context of this study with uncertain renewable wind generation
dependent on a stochastic space of \emph{iid} standard normal random
variables, $\xivec=\{\xi_1,\xi_2,\ldots,\xi_n\}$,
we employ Hermite polynomials to construct a Hermite-Gauss (HG) PCE
for the (minimum) cost $Q(\bm{x},\xivec)$. We employ Eq.~(\ref{eq:pce}) for a
truncated HG PCE
\begin{equation}
Q_\mathrm{PC}(\xvec,\xivec) = \sum_{k=0}^P c_k(\xvec)\Psi_k(\xivec),
\label{eq:qpce}
\end{equation}
where $\Psi_k(\xivec)$ are $n$-variate Hermite polynomials.
The coefficients $c_k$ depend on the discrete variable $\bm{x}$, hence
separate PCE approximations for $Q$ will be constructed for each
instance of $\bm{x}$ chosen in the unit commitment stage. Given Eq.~(\ref{eq:proj}), we have
\be
c_k(\bm{x}) = \frac{\langle Q\Psi_k\rangle}{\langle \Psi_k^2\rangle}=
\frac{1}{\langle \Psi_k^2\rangle}\int_{\mathfrak{R}^n}
Q(\bm{x},\xivec)\Psi_k(\xivec) p(\xivec)\,d\xivec.
\label{eq:gproj}
\ee
where we have used $p(\xivec)$ as a product of univariate standard
normal PDFs. The dimensionality of the PCE in Eq.~\eqref{eq:qpce} is
the same as the number of stochastic dimensions used to represent the
uncertain renewable power. For the study presented in this paper,
$n=16$.

Given $Q_\mathrm{PC}(\xvec,\xivec)$,
then, we have
\be
{\overline Q}(\bm{x})=\mathrm{E}_\xivec[Q(\bm{x},\xivec)]=\langle
Q(\bm{x},\xivec) \rangle = c_0,
\ee
being the solution of the stochastic ED problem. The PCE in
Eq.~\eqref{eq:qpce} can also be used to generated higher-order moments
for the minimum cost, based directly on the coefficients for the
corresponding PC basis terms.

Several methods can be employed to evaluate the projection integrals in
Eq.~(\ref{eq:gproj}). MC methods can be used in principle, but are impractical
given their slow convergence rate. Alternatively, for smooth
integrands, and particularly in low-moderate dimensional problems, sparse
quadrature methods~\cite{Smolyak:1963,Gerstner:1998,Conrad:2013} can provide
highly accurate results with smaller numbers of deterministic samples.
\begin{figure}[h!]
  \centering
    \includegraphics[width=0.23\textwidth]{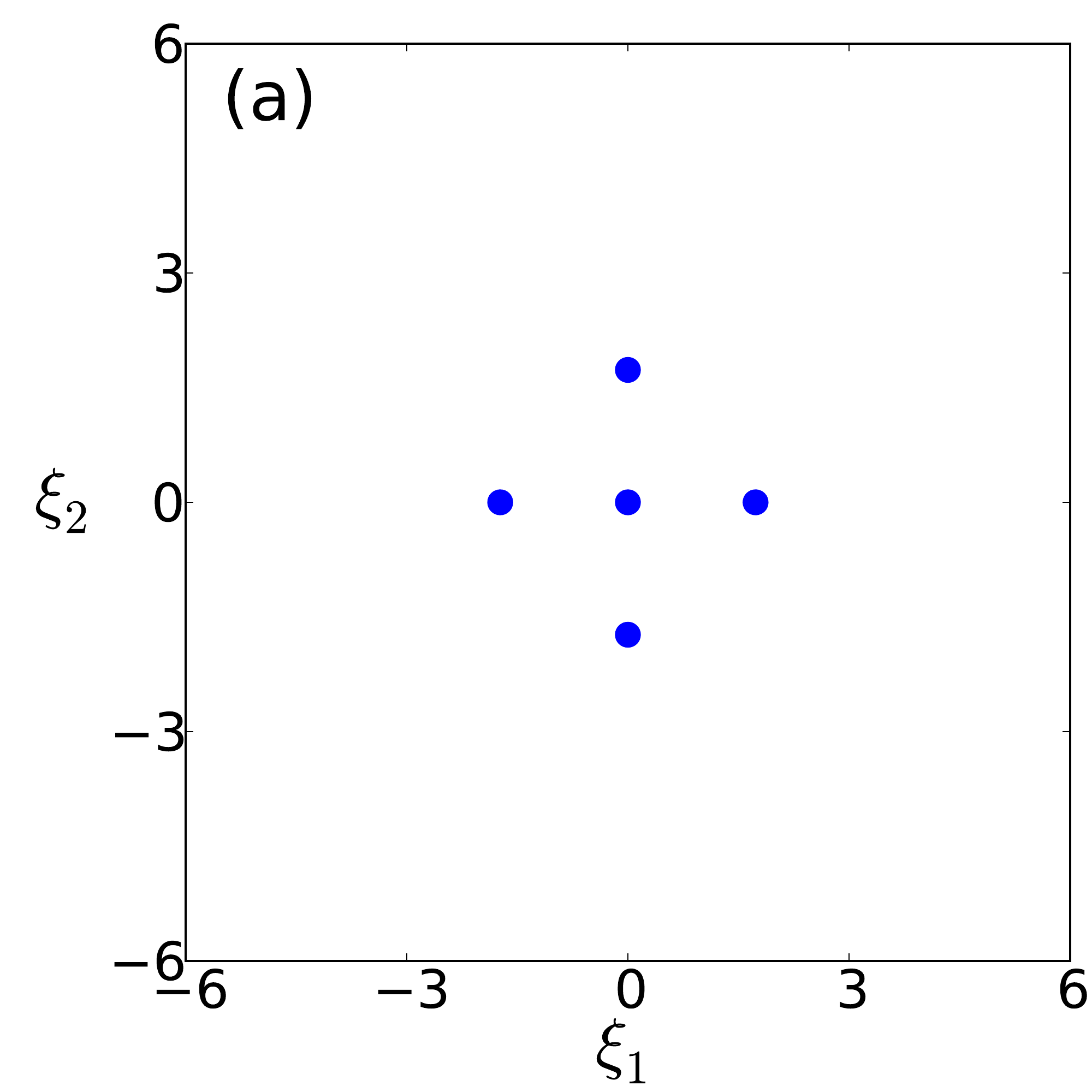}
    \includegraphics[width=0.23\textwidth]{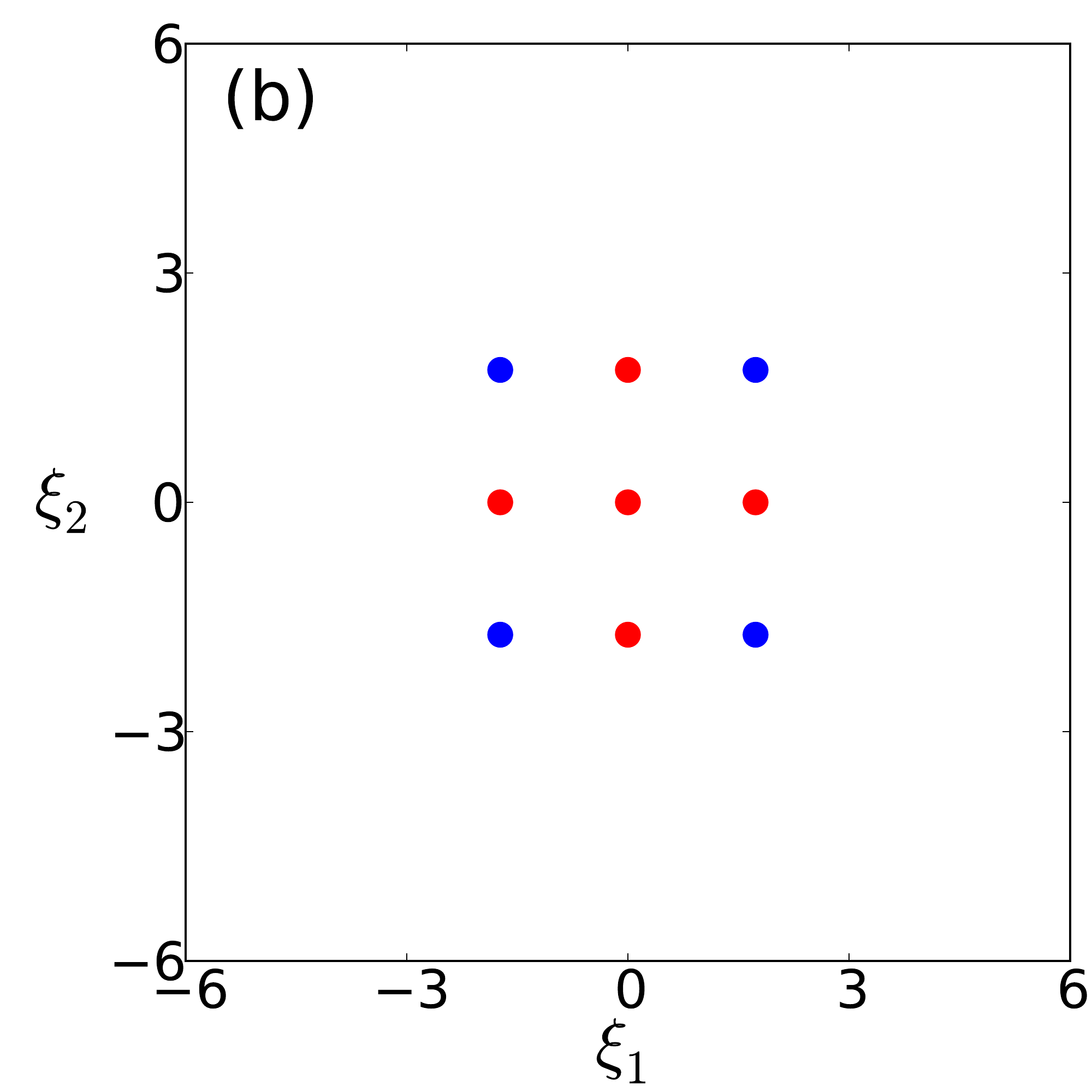}
    \includegraphics[width=0.23\textwidth]{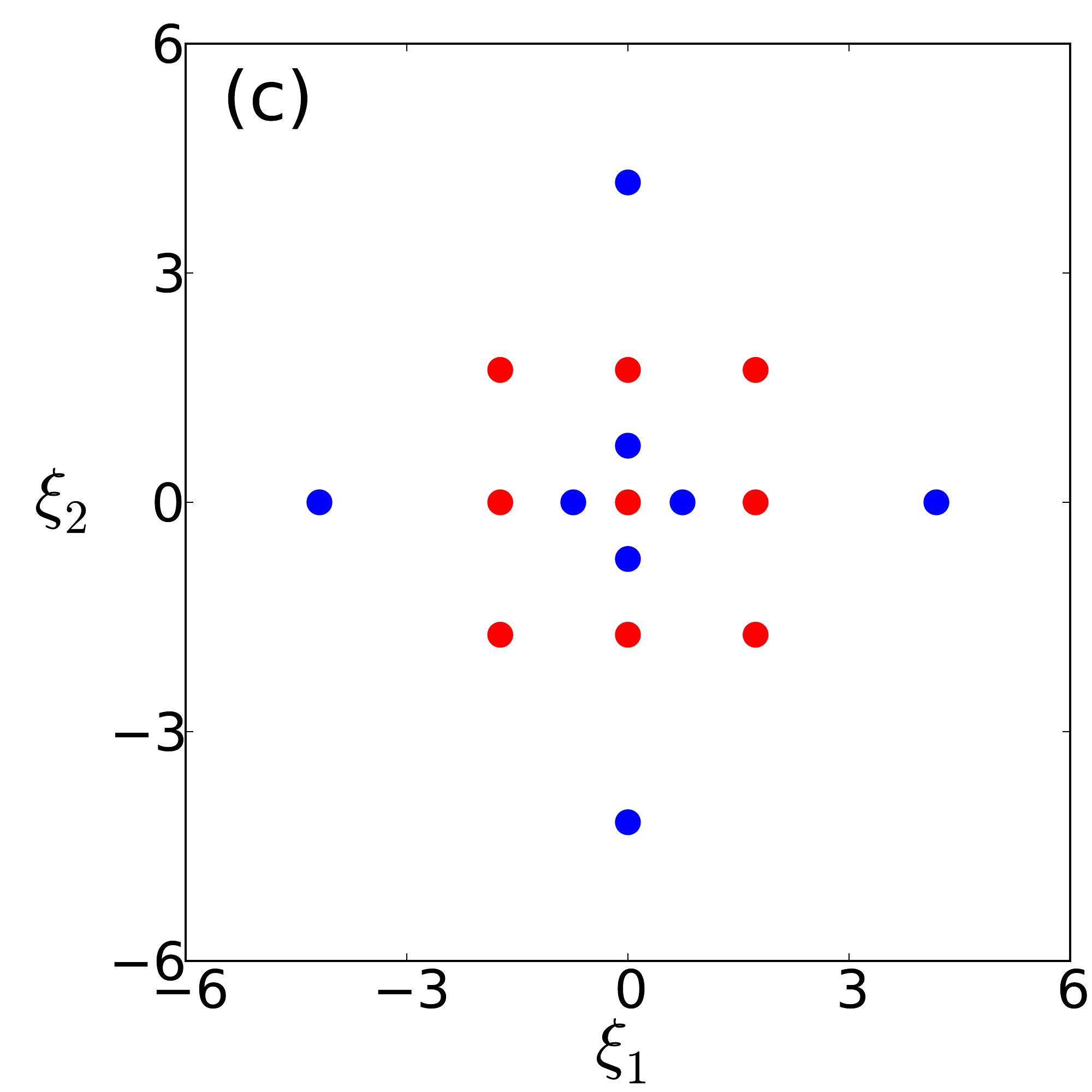}
    \includegraphics[width=0.23\textwidth]{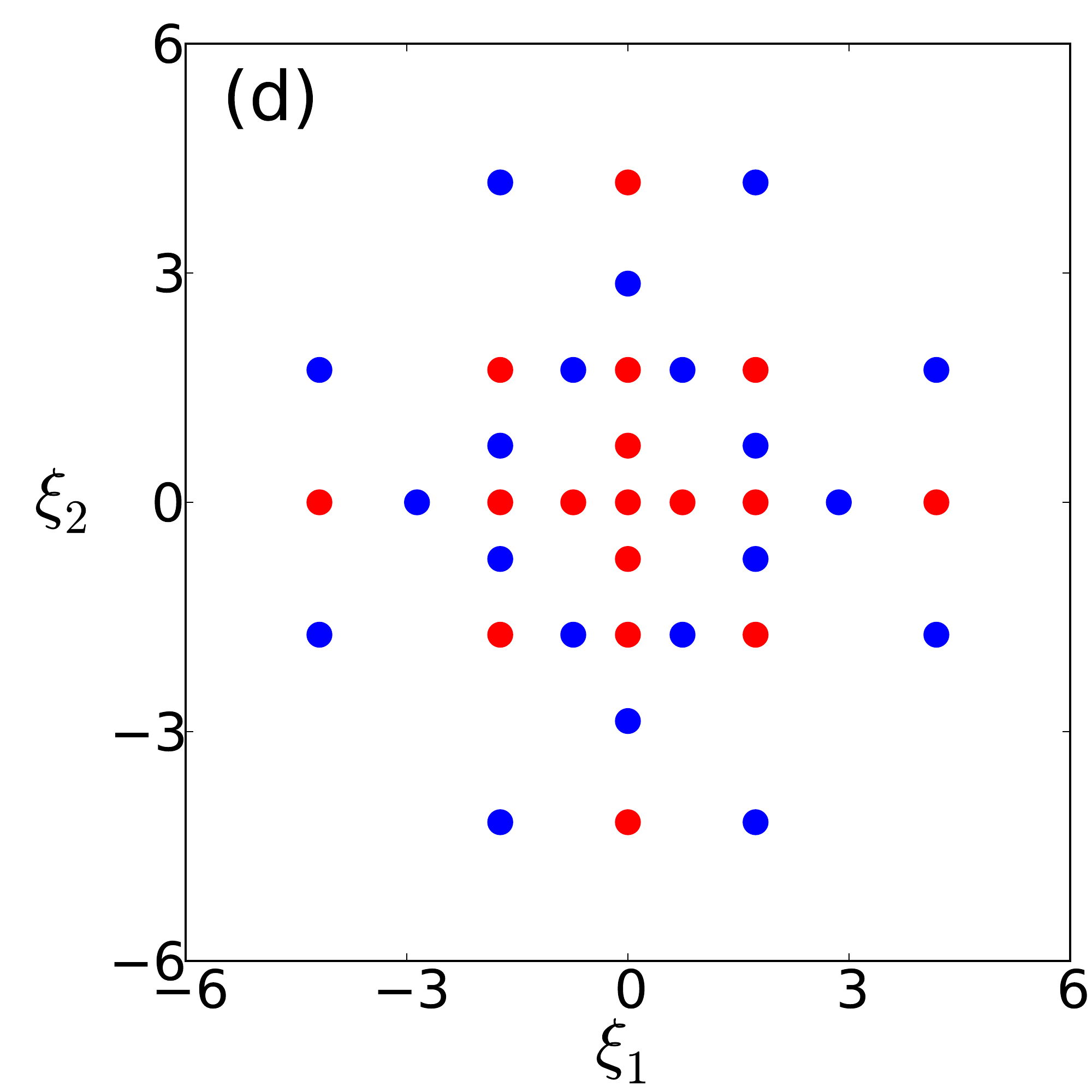}
  \caption{Placement of deterministic samples via a sparse grid
    approach: (a) Level 1, (b) Level 2, (c) Level
    3, and (d) Level 4. For Levels 2 through 4, the red symbols show
    samples for previous levels while blue symbols show newly added samples.
    \label{fig:sgr}}
\end{figure}
Fig.~\ref{fig:sgr} shows, in a 2D configuration, the locations of
deterministic samples we use, with a sparse grid employing
Gauss-Kronrod quadrature~\cite{Patterson:1968}. Several levels are shown
in the figure, starting with the first level in Fig.~\ref{fig:sgr}(a),
and following with additional samples leading to Levels 2, 3, and 4,
in Fig.~\ref{fig:sgr}(b)-(d). A first order PCE requires a Level 2
sparse grid, while a second order PCE requires a Level 3 grid.
The number of requisite samples using sparse-quadrature evaluation of
the projection integrals, for a given requisite surrogate accuracy, is
much smaller than the corresponding number of MC samples, as we will
illustrate in the next subsection.

\subsection{Numerical Results}
\label{sec:results}

We now present numerical results comparing the expected costs of our SED
model using scenarios obtained with both our Polynomial Chaos approach to
sampling versus a traditional Monte Carlo approach. We consider
the IEEE 118-bus test system~\cite{IEEEdata} augmented with the three
wind generation sites discussed in Section~\ref{sec:windkle}.
The three renewable generators at sites 15414, 16238, and 3560 replace the
conventional generators at buses 89, 69, and 10, respectively.
We employ $|T|=24$ time periods and represent stochastic renewable generation
via the Karhunen-Loeve expansions presented in Section~\ref{sec:windkle}. We
use 6 KL modes  per site. Given the strong dependence between the first two RVs
for sites 15414 and 16238, the effective dimensionality of the system decreases
by 2, to a total of 16.

\begin{figure}[t!]
  \centering
    \includegraphics[width=0.23\textwidth]{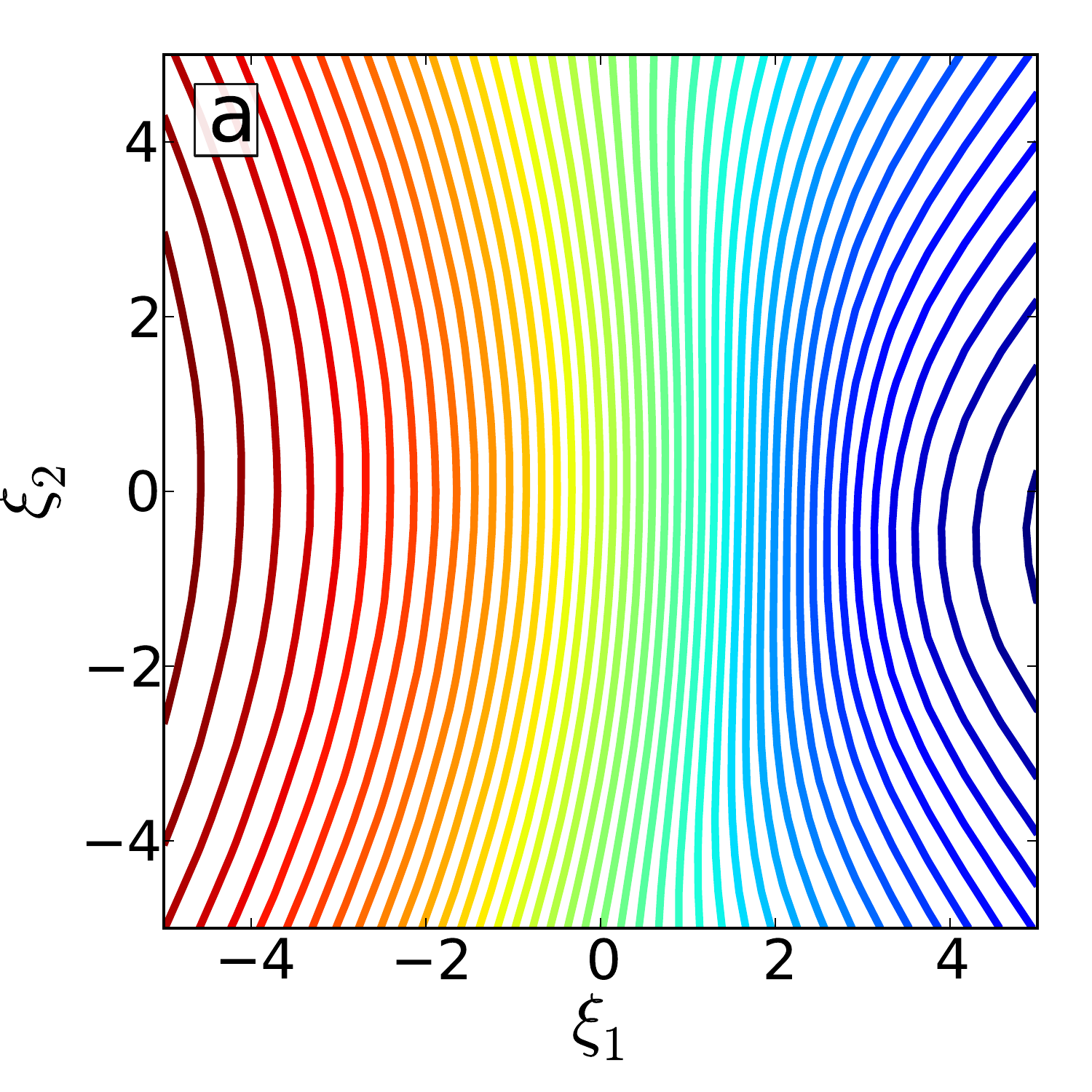}
    \includegraphics[width=0.23\textwidth]{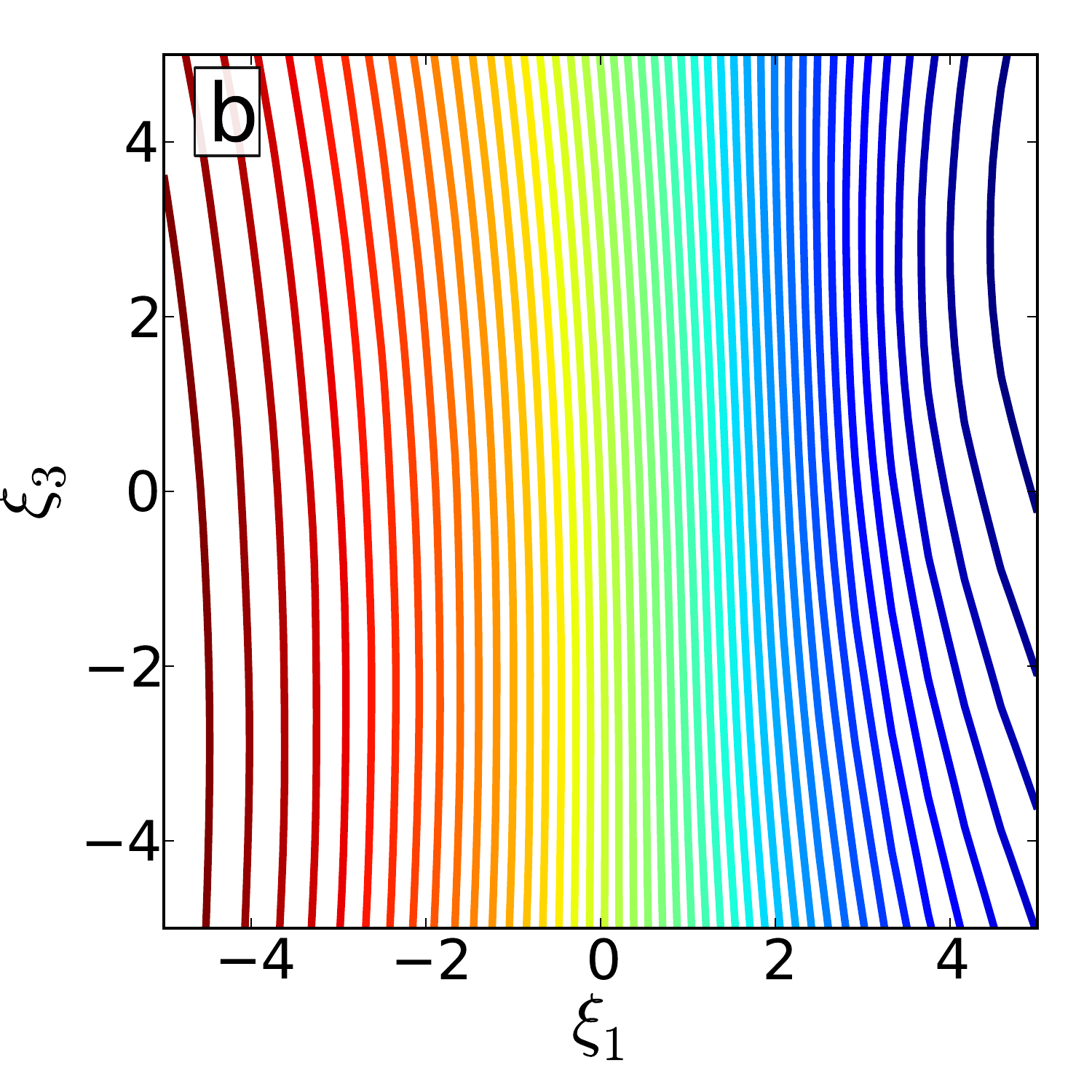}
    \includegraphics[width=0.23\textwidth]{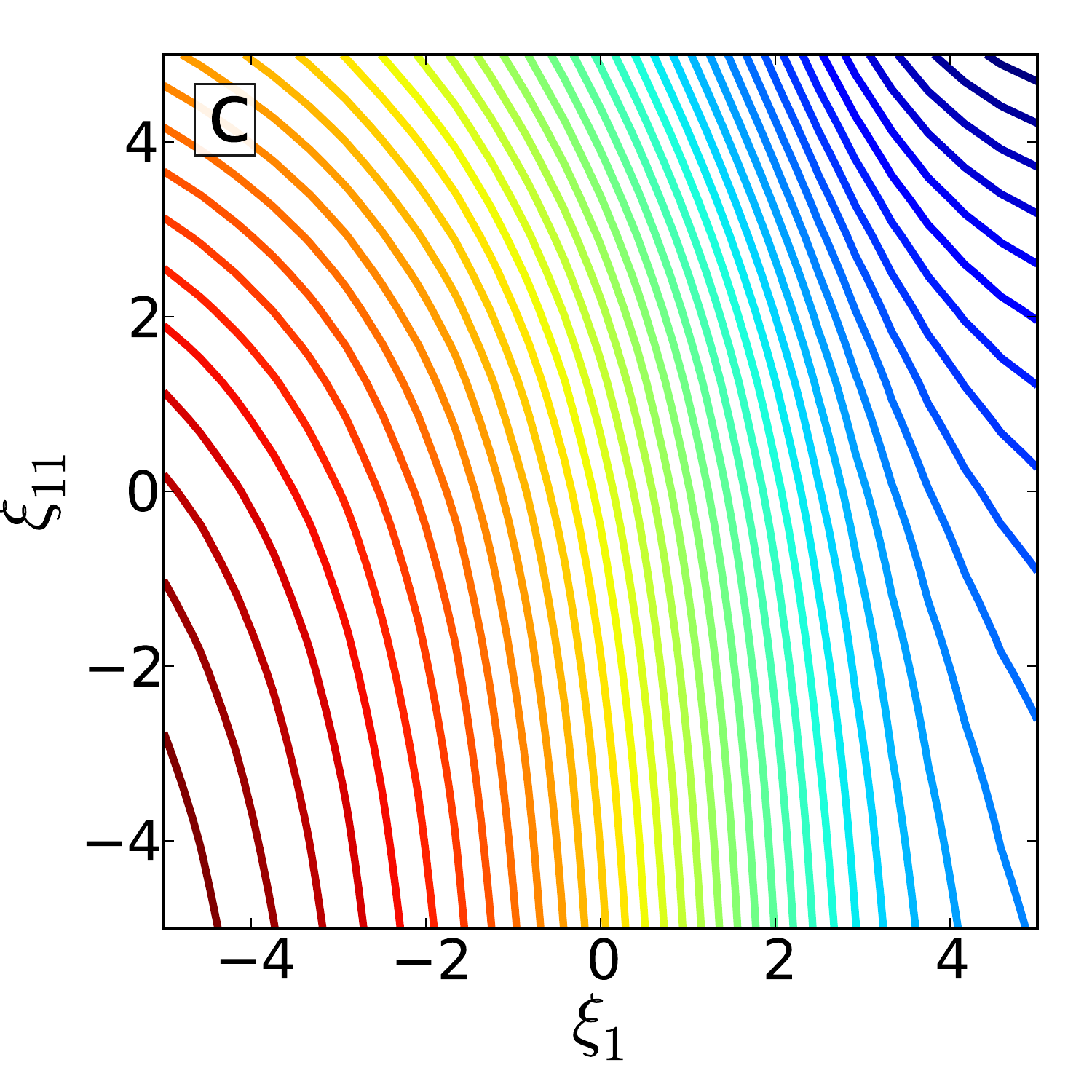}
    \includegraphics[width=0.23\textwidth]{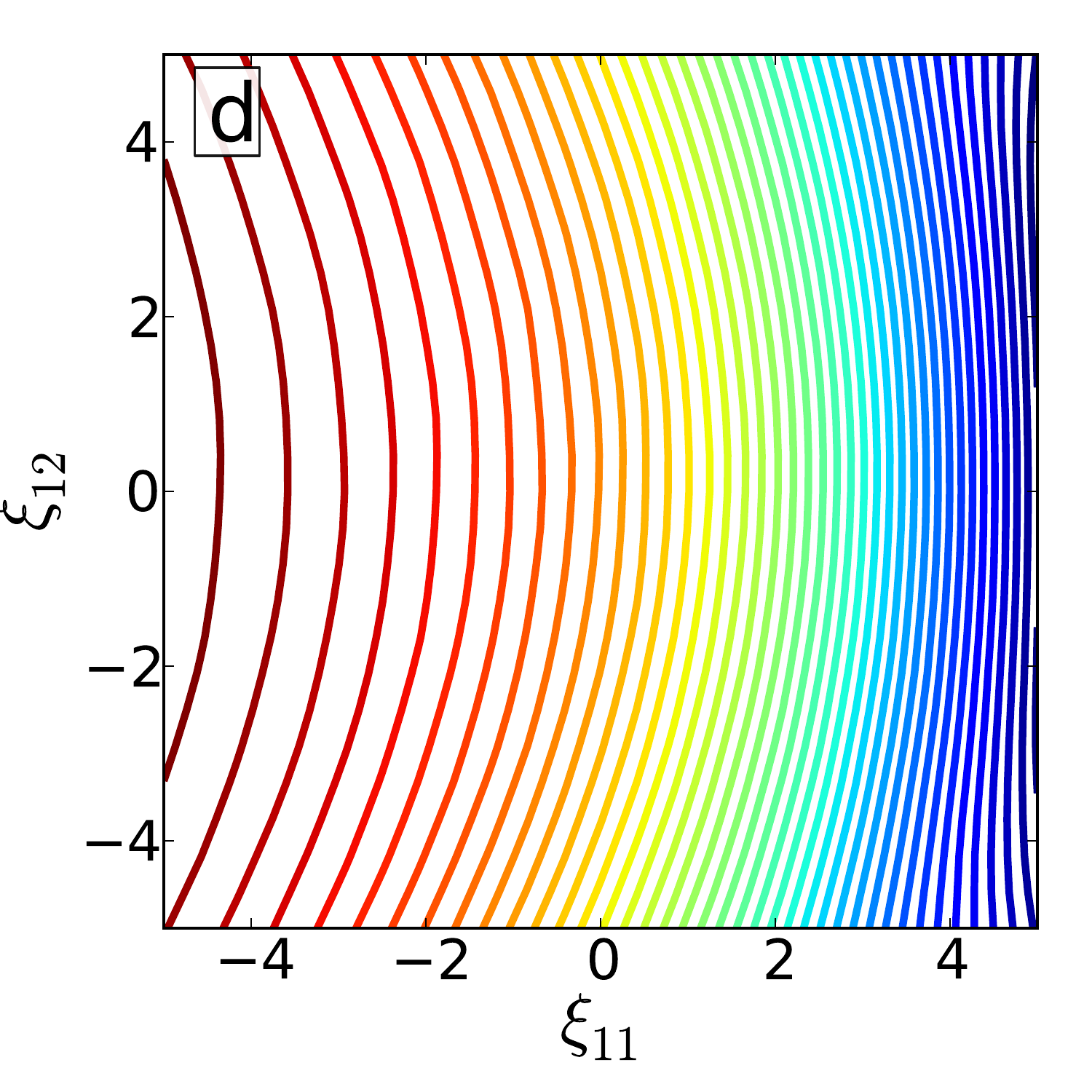}
  \caption{Two dimensional slices through the center of the
    stochastic space, $\xivec=0$. Random variables corresponding to (a)
    first two modes for Wyoming sites, (b) 1st and 3rd mode for WY sites,
    (c) first modes for WY and CA sites, and (d) first two modes for CA site.
    The countour lines correspond to iso-levels of $Q$, increasing from
    blue to yellow to red.
    \label{fig:sl2D}}
\end{figure}

We first explore the dependence of $Q(\xvec,\xivec)$ on the stochastic space
that characterizes renewable generation. Fig.~\ref{fig:sl2D} shows 2D slices
through the 16D $\xivec$ space. All slices are anchored at the origin
$\xivec=0$. RVs $\xi_1$ through $\xi_6$ correspond to site 16238, while
$\xi_1$,  $\xi_2$, and $\xi_7$ - $\xi_{10}$ correspond to site 15414. The
remaining random variables, $\xi_{11}$ - $\xi_{16}$, correspond to site 3560.
The expected cost is about $\$2.4$ million, and the relative change about the
mean is about $15\%$ in the numerical tests presented here.
The results shown in the top row of Fig.~\ref{fig:sl2D} indicate that $Q$ is
strongly dependent on $\xi_1$, while the dependence on $\xi_2$ and $\xi_3$
is weaker -- as suggested by negligible contour changes in those directions.
The slice in the $(\xi_1,\xi_{11})$ plane highlights the contribution of $\xi_{11}$
to the variation of $Q$, while the slice in the $(\xi_{11},\xi_{12})$ plane indicates
less impact of the second mode for site 3560 on cost. Other 2D slices (not shown)
confirm the diminishing impact that higher-order modes have on $Q$.
Collectively, these results also indicate that $Q(\xvec,\xivec)$ is smooth in
$\xivec$, which makes it amenable to a PCE representation.

\begin{figure}[t!]
  \centering
    \includegraphics[width=0.45\textwidth]{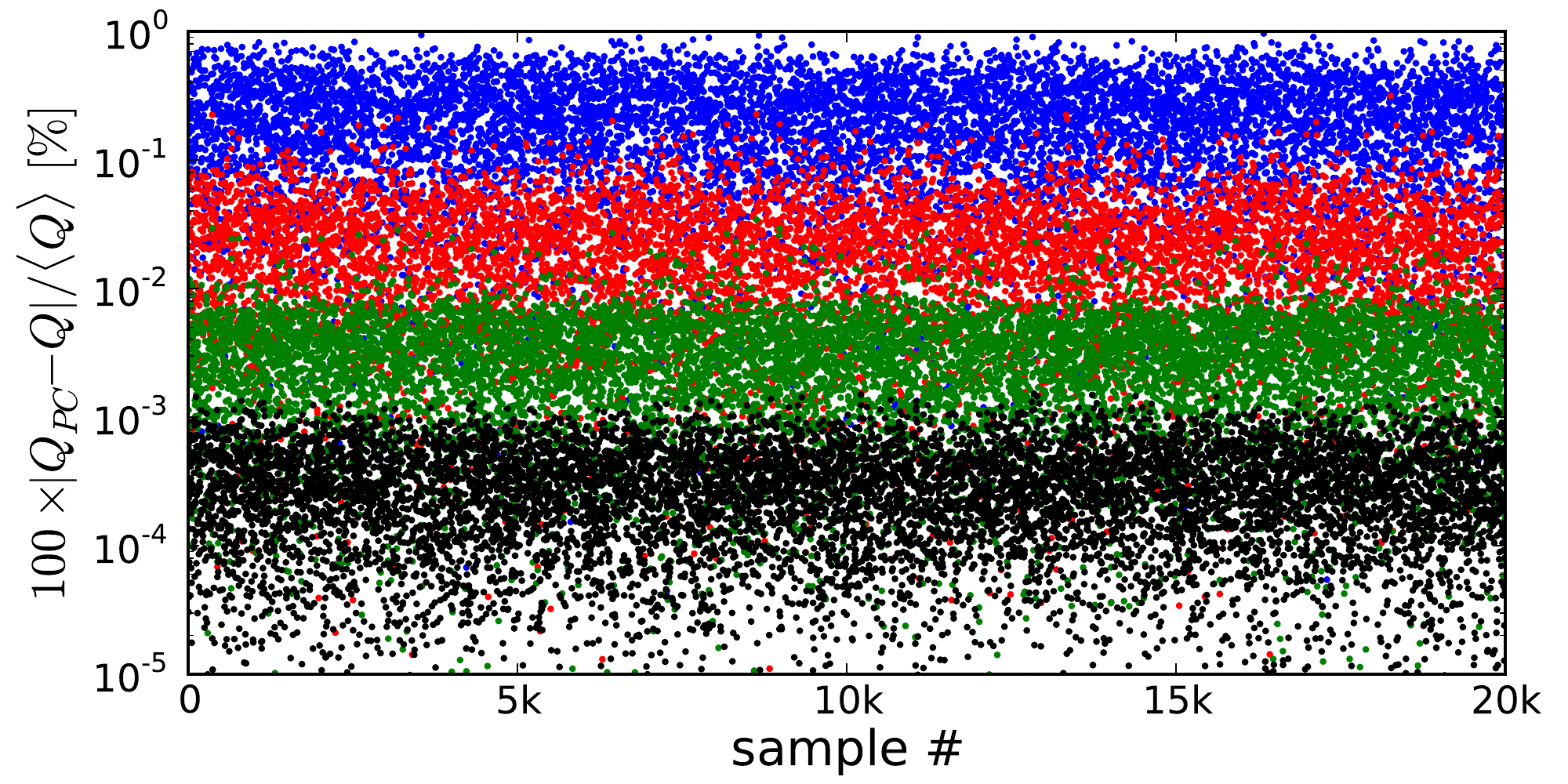}
  \caption{Cross-validation of PCE approximation for the minimum cost
    $Q(\xvec,\xivec)$. The blue, red, green, and black symbols correspond to
    1st through 4th order PCEs.\label{fig:errQ}}
\end{figure}

We next proceed to test the accuracy of the PCE representation for
$Q(\xvec,\xivec)$ with respect to actual model evaluations. We construct
several PCEs, from 1st to 4th order. Given the 16-dimensional stochastic
space corresponding to the three wind sites, the number of sparse
quadrature sample points necessary for the construction of the PCE
coefficients is approximately $5\times 10^2$ for a 1st order expansion,
$5\times 10^3$ for a 2nd order expansion, $3.6\times 10^4$ for a
3rd order expansion, and $2\times 10^5$ for a 4th order expansion.
We cross-validate our PCE representations relative to exact solutions
at $5\times 10^5$ randomly chosen $\xivec$ samples. Fig.~\ref{fig:errQ}
shows the relative $L_1$ error between actual model evaluations and the
PCE-approximated cost. This error is relative to the expected minimum
cost computed using all model evaluations available and is converted to
percentages. For brevity, only a random subset of $2\times 10^4$ samples are shown.
These results indicate a drop of about one order of magnitude, per
polynomial degree, for the magnitude of the error between full model evaluations
and the PCE results.

The above analysis explores the dependence of minimum cost for individual
samples on the stochastic space corresponding to the uncertain wind at select
sites, and the accuracy of the PCE approach in capturing this dependence.
However, in the context of the SED problem, and for the present work
in particular, we are only interested the accuracy of the expected minimum cost
$\left<Q(\xvec,\xivec)\right>$.\footnote{Our approach is not restricted to expected
costs; other moments of $Q$ can be similarly considered.} Hence, our goal here
is to demonstrate the efficiency of a quadrature approach to estimate the expectation.

\begin{figure}[h!]
  \centering
    \includegraphics[width=0.45\textwidth]{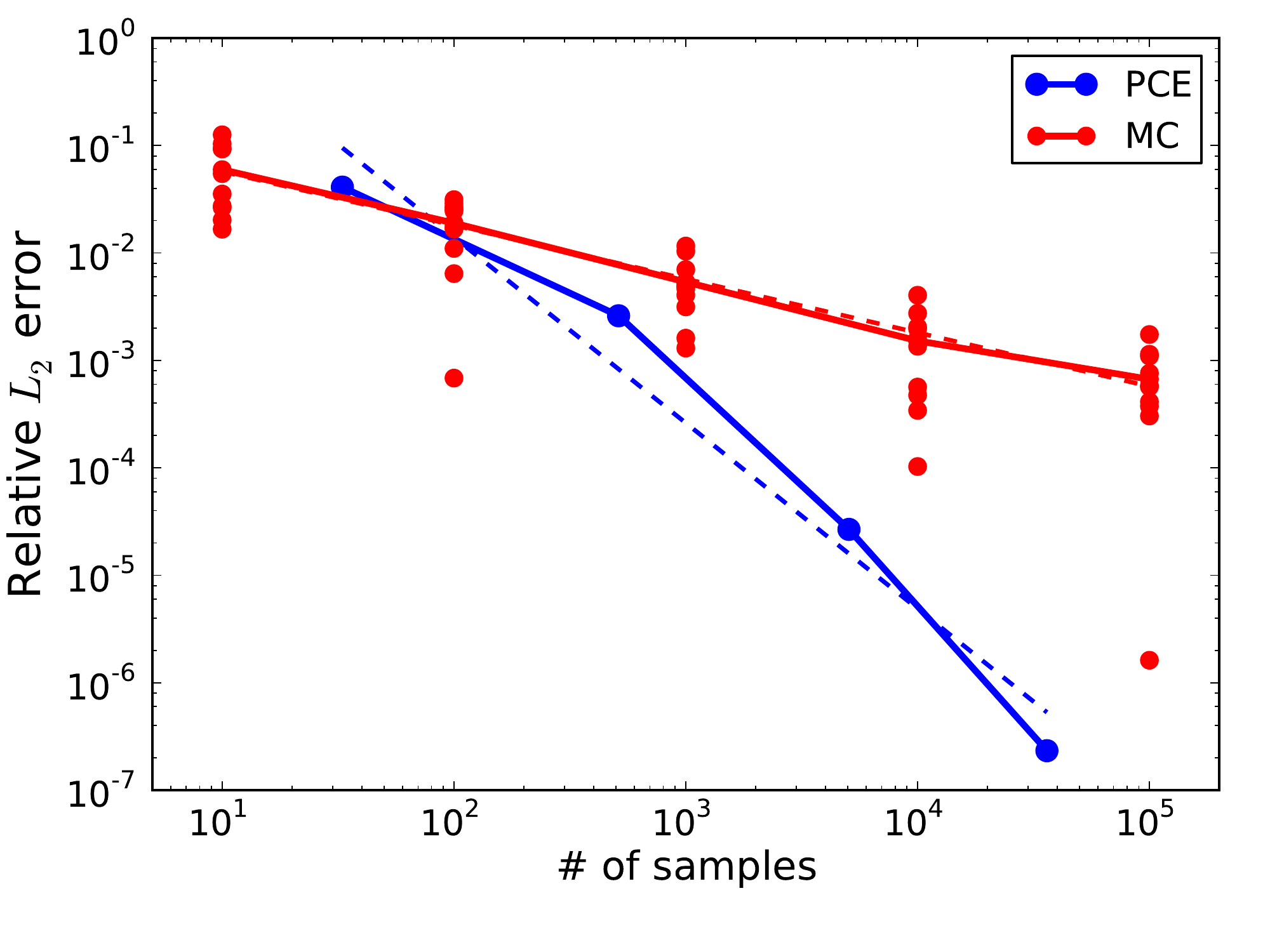}
  \caption{Comparison of errors in the estimation of expected cost
    between PCE-based results and Monte Carlo estimates. Dashed lines
    show power-law fits, $a\times N^b$ to test the convergence rates of the MC
    and PCE approaches.\label{fig:econv}}
\end{figure}

Toward this goal, we show the convergence of $\left<Q(\xvec,\xivec)\right>$ as
a function of the number of samples considered. For the PCE approach, the expected
minimum cost is simply given by the $c_0$ coefficient in Eq.~\eqref{eq:gproj}. The
results shown with blue are based on this coefficient, computed with increasingly
accurate sparse quadratures. The error measure is relative, being normalized with
respect to the value corresponding to the next higher accuracy result, defined
as
\begin{equation}
E_{PC,i}=\frac{\vert {{c_0}}_{i}-{{c_0}}_{i+1}\vert}{{{c_0}}_{i+1}}
\end{equation}
where the subscript $i$ denotes the quadrature level, e.g., $33$ model evaluations
for level $1$, $513$ model evaluations for level $2$, and so forth. The error for the MC
results is computed using a similar approach. However, because results for the
MC approach depend on a randomly drawn set of samples, we show results for
several realizations. The specific formula is given as
\begin{equation}
E_{MC,i}^j=\frac{\vert \overline{Q}_{i}^j-\overline{\overline{Q}}_{i+1}\vert}
{\overline{\overline{Q}}_{i+1}}
\end{equation}
where the subscript $i$ denotes the number of samples used to compute the
average cost $\overline{Q}_{i}^j$, e.g., 10 samples for $i=1$, 100 samples for $i=2$,
etc. We employed 10 realizations, indexed by superscript $j$, for each set of
MC samples; the average cost over all realizations is denoted using a double
overline. $E_{MC,i}^j$ is shown with red circles in Fig.~\ref{fig:econv}, while
the red line connects the mean error over all realizations for a particular
number of MC samples. The dashed lines show power-law curve fits, $a\times N^{-b}$,
for the data, which we use to analyze convergence rates as a function of the
number of samples. For the MC approach we recover the theoretical convergence
rate, $b=0.5$, while for the PCE approach $b=1.9$. Visual inspection of the PCE
results and power law fits indicate a stronger polynomial convergence rate than
MC, rather than an exponential rate (results not shown) corresponding to spectral
accuracy.

In this figure a $1\%$ error level corresponds to about $\$24$K, while
a $0.01\%$ error level is in the hundreds of dollars. These results
indicate that if a low degree of accuracy,
\emph{e.g.} $1\%$, is sufficient or desired, then the PCE
approach exhibits a computational cost that is comparable to the
MC approach. Nevertheless there is a considerable spread in the
accuracy of these estimates, making MC estimates with a small number of samples
somewhat unreliable. For situations where higher accuracies are
required, for example of $0.01\%$, the PCE estimates require one to
two orders of magnitude less model evaluations compared to the MC
estimates.

\section{Conclusion}
\label{sec:conclusion}

In this paper, we present methods for efficient representation of
uncertainty, with emphasis on the Stochastic Economic
Dispatch problem. We present two, related, category of methods, one
for spectral random fields representations, and one for functional
representation of random variables.

In the first part of this paper we employ data on renewable wind
generations provided by NREL. We determine that 24-hour wind
samples can be represented via Karhunen-Loeve (KL) expansions. This
expansion represents stochastic processes by a linear combination of
orthogonal modes. Further, the KL representation employs the basis
that provides the minimum total $L_2$ error. Analysis of the KL
eigenspectra at several wind sites indicate that $6$ KL modes are
sufficient to capture about $95\%$ of the total variance, effectively
reducing the dimensionality of the stochastic space by a factor of
$4$. The dimensionality of the stochastic space is further reduced by
exploiting the dependence between KL random variables for wind sites
that are geographically close.

In the second part of the paper we present an approach to reduce the
computational cost associated with stochastic unit commitment and
economic dispatch, by reducing the number of required forecast
samples. This approach is based on Polynomial Chaos expansion (PCE)
models for the production cost that cover the uncertainty in the
renewable generation. The construction of the PCE terms is based on
the projection of the model on increasingly higher basis
modes. Consequently, the global error in an $L_2$ sense between the
surrogate model and the actual simulations is easily controlled.

We present computational results for the 118-bus test case, augmented
to account for renewable generation. We find that quadratic PCE models
for the production cost showed pointwise relative errors less than
$1\%$ throughout the uncertain demand space. For relative accuracies
of $o(10^{-4})$, the PCE approach is one to two orders of
computationally more efficient compared to Monte Carlo estimates.

We plan to extend the methods presented here to higher-dimensional
power grid problems. In order to alleviate the curse of dimensionality
in such studies we plan to emply global sensitivity analysis to
eliminate the stochastic variables that are not important for the
expected cost. And based on results observed in this paper, we plan
to explore adaptive sparse quadrature constructions to tailor of the
PCE order to the specific dependence of minimum cost on each
component of the stochastic space.

\ifCLASSOPTIONcompsoc
  \section*{Acknowledgments}
\else
  \section*{Acknowledgment}
\fi
This work was funded by the Laboratory Directed Research \&
Development (LDRD) program at Sandia National Laboratories. Sandia
National Laboratories is a multiprogram laboratory operated by Sandia
Corporation, a wholly owned subsidiary of Lockheed Martin Corporation,
for the United States Department of Energy's National Nuclear Security
Administration under contract DE-AC04-94AL85000.

\ifCLASSOPTIONcaptionsoff
  \newpage
\fi

\clearpage
\section*{Appendix. Total Variance of Karhunen-Loeve Expansion}

Starting from Eq.~\eqref{eq:kle}, it follows that
\[
\textrm{Var}[W_L(t,\omega)]=\textrm{Var}\left[\sum_{k=1}^{24}
\sqrt{\lambda_k} f_k(t) \xi_k\right]
\]
Here, we converted the infinite sum into one over the $24$ modes that
can be extracted from 24h wind samples. As noted in
Section~\ref{sec:windkle}, $\xi_k$ are uncorrelated random variables
with zero mean and unit variance. To simplify notation, in this
Appendix, we change notation and denote the product $\sqrt{\lambda_k}
\xi_k$ as $X_k$. This new set of RVs are uncorrelated with zero mean
and variance $\lambda_k$.

The variance of the {\it rhs} of the expression above can be written
as\hfill

\begin{strip}
\begin{eqnarray}
\textrm{Var}\left[\sum_{k} f_k(t)
X_k\right]&=&\mathbb{E}\left[\left(\sum_{k} f_k(t)
  X_k-\mathbb{E}\left[\sum_{k} f_k(t) X_k\right]\right)^2\right]
=\mathbb{E}\left[\left(\sum_{k} f_k(t)
  \left(X_k-\mathbb{E}\left[X_k\right]\right)\right)^2\right] \\
&=&\mathbb{E}\left[\sum_{k} f_k^2(t)
    \left(X_k-\mathbb{E}\left[X_k\right]\right)^2+2\sum_{i\neq
      j}f_i(t)f_j(t) \left(X_i-\mathbb{E}\left[X_i\right]\right)
    \left(X_j-\mathbb{E}\left[X_j\right]\right)\right]\\
&=&\sum_{k} f_k^2(t)
\underbrace{\mathbb{E}\left[\left(X_k-\mathbb{E}\left[X_k\right]\right)^2\right]}_{\textrm{Var}[X_k]}
+2\sum_{i\neq
      j}f_i(t)f_j(t) \mathbb{E}\left[\left(X_i-\mathbb{E}\left[X_i\right]\right)
    \left(X_j-\mathbb{E}\left[X_j\right]\right)\right]\label{eq:vsum2}
\end{eqnarray}
\end{strip}
For uncorrelated random variables $X_i$, $X_j$, the expectation in the
second sum in Eq.~\eqref{eq:vsum2} is identically zero,
$\mathbb{E}\left[\left(X_i-\mathbb{E}\left[X_i\right]\right)
    \left(X_j-\mathbb{E}\left[X_j\right]\right)\right]=0$. Making use
  of $\textrm{Var}[X_k]=\lambda_k$, it follows that
\be
\textrm{Var}\left[\sum_{k} f_k(t)
X_k\right]=\sum_{k} f_k^2(t)\textrm{Var}[X_k] =\sum_{k} f_k^2(t)\lambda_k
\ee
The total variance over the entire deterministic space is given by
\[
\int_t \textrm{Var}\left[\sum_{k} f_k(t)
X_k\right] dt=\int_t \sum_{k} f_k^2(t)\lambda_k dt
\]
Given that modes $f_k$ are orthonormal over the deterministic space,
total variance of the KL expansion is given by
\[
\int_t \textrm{Var}\left[W_L(t,\omega)\right] dt=\sum_{k} \lambda_k
\]


\begin{thebibliography}{10}
\providecommand{\url}[1]{#1}
\csname url@samestyle\endcsname
\providecommand{\newblock}{\relax}
\providecommand{\bibinfo}[2]{#2}
\providecommand{\BIBentrySTDinterwordspacing}{\spaceskip=0pt\relax}
\providecommand{\BIBentryALTinterwordstretchfactor}{4}
\providecommand{\BIBentryALTinterwordspacing}{\spaceskip=\fontdimen2\font plus
\BIBentryALTinterwordstretchfactor\fontdimen3\font minus
  \fontdimen4\font\relax}
\providecommand{\BIBforeignlanguage}[2]{{%
\expandafter\ifx\csname l@#1\endcsname\relax
\typeout{** WARNING: IEEEtran.bst: No hyphenation pattern has been}%
\typeout{** loaded for the language `#1'. Using the pattern for}%
\typeout{** the default language instead.}%
\else
\language=\csname l@#1\endcsname
\fi
#2}}
\providecommand{\BIBdecl}{\relax}
\BIBdecl

\bibitem{Carrion:2006}
M.~Carri\'{o}n and J.~M. Arroyo, ``A computationally efficient mixed-integer
  linear formulation for the thermal unit commitment problem,'' \emph{{IEEE
  Transactions on Power Systems}}, vol.~21, no.~3, pp. 1371--1378, 2006.

\bibitem{isone:2014}
``{ISO} {N}ew {E}ngland: Forecast and scheduling reserve adequacy
  analysis\hfill,'' \url{www.iso-ne.com/support/training/courses/wem101/
  \linebreak 10\_forecast\_scheduling\_callan.pdf}, accessed: 2014-05-19.

\bibitem{Ruiz:2009}
P.~A. Ruiz, R.~C. Philbrick, E.~Zack, K.~W. Cheung, and P.~W. Sauer,
  ``Uncertainty management in the unit commitment problem,'' \emph{{IEEE
  Transactions on Power Systems}}, vol.~24, no.~2, pp. 642--651, 2009.

\bibitem{Takriti:96}
S.~Takriti, J.~Birge, and E.~Long, ``A stochastic model for the unit commitment
  problem,'' \emph{{IEEE} Transactions on Power Systems}, vol.~11, no.~3, pp.
  1497--1508, 1996.

\bibitem{Papavasiliou:2013}
A.~Papavasiliou and S.~S. Oren, ``{Multiarea Stochastic Unit Commitment for
  High Wind Penetration in a Transmission Constrained Network},''
  \emph{Operations Research}, vol.~61, no.~3, pp. 578--592, 2013.

\bibitem{Wang:2013}
Q.~Wang, J.~Wang, and Y.~Guan, ``Stochastic unit commitment with uncertain
  demand response,'' \emph{{IEEE Transactions on Power Systems}}, vol.~28,
  no.~1, pp. 562--563, 2013.

\bibitem{Chen:2014}
R.~L.-Y. Chen, N.~Fan, A.~Pinar, and J.-P. Watson, ``Contingency-constrained
  unit commitment with post-contingency corrective recourse,'' \emph{Annals of
  Operations Research}, pp. 1--27, 2014.

\bibitem{Bertsimas:2013}
D.~Bertsimas, E.~Litvinov, X.~A. Sun, J.~Zhao, and T.~Zheng, ``Adaptive robust
  optimization for the security constrained unit commitment problem,''
  \emph{{IEEE Transactions on Power Systems}}, vol.~28, no.~1, pp. 52--63,
  2013.

\bibitem{Thiam:2010}
F.~B. Thiam and C.~L. DeMarco, ``Optimal transmission expansion via intrinsic
  properties of power flow conditioning,'' in \emph{2010 North American Power
  Symposium (NAPS)}, 2010, pp. 1--8.

\bibitem{Najm:2009a}
H.~N. Najm, ``{Uncertainty Quantification and Polynomial Chaos Techniques in
  Computational Fluid Dynamics},'' \emph{Annual Review of Fluid Mechanics},
  vol.~41, no.~1, pp. 35--52, 2009.

\bibitem{Ghanem:1991}
R.~G. Ghanem and P.~D. Spanos, \emph{Stochastic Finite Elements: A Spectral
  Approach}.\hskip 1em plus 0.5em minus 0.4em\relax Springer Verlag, New York,
  1991.

\bibitem{Safta:2014}
C.~Safta, R.~L.-Y. Chen, H.~Najm, A.~Pinar, and J.-P. Watson, ``Toward using
  surrogates to accelerate solution of stochastic electricity grid operations
  problems,'' in \emph{North American Power Symposium (NAPS), 2014}, Sept 2014,
  pp. 1--6.

\bibitem{OlmOmk:2010}
O.~{Le Ma{\^\i}tre} and O.~Knio, \emph{{Spectral Methods for Uncertainty
  Quantification}}.\hskip 1em plus 0.5em minus 0.4em\relax New York, NY:
  Springer, 2010.

\bibitem{nrelwwd:2015}
``{NREL: Transmission Grid Integration - Western Wind Dataset},''
  \url{http://www.nrel.gov/electricity/transmission/wind\_integration\_dataset.html},
  accessed: 2015-02-28.

\bibitem{Nystrom:1930}
E.~J. Nystr\"{o}m, ``\"{U}ber die praktische aufl\"{o}sung von
  integralgleichungen mit anwendungen auf randwertaufgaben,'' \emph{Acta
  Mathematica}, vol.~54, no.~1, pp. 185--204, 1930.

\bibitem{Szekely:2007}
G.~J. Sz\'{e}kely, M.~L. Rizzo, and N.~K. Bakirov, ``Measuring and testing
  dependence by correlation of distances,'' \emph{Annals of Statistics},
  vol.~35, pp. 2769--2794, 2007.

\bibitem{Safta:2014b}
C.~Safta, K.~Sargsyan, H.~N. Najm, K.~Chowdhary, B.~Debusschere, L.~P. Swiler,
  and M.~S. Eldred, ``{Probabilistic Methods for Sensitivity Analysis and
  Calibration of Computer Models in the NASA Challenge Problem},''
  \emph{Journal of Aerospace Information Systems}, 2015, in press.

\bibitem{Rasmussen:2006}
C.~E. Rasmussen and C.~K.~I. Williams, \emph{Gaussian Processes for Machine
  Learning}.\hskip 1em plus 0.5em minus 0.4em\relax {MIT} {P}ress, 2006.

\bibitem{eliawwd:2015}
``{ELIA: Wind-Power Generation Data},''
  \url{http://www.elia.be/en/grid-data/power-generation/wind-power}, accessed:
  2015-07-26.

\bibitem{Wiener:1938}
N.~Wiener, ``The homogeneous chaos,'' \emph{Am. J. Math.}, vol.~60, pp.
  897--936, 1938.

\bibitem{Janson:1997}
S.~Janson, \emph{{Gaussian Hilbert Spaces}}.\hskip 1em plus 0.5em minus
  0.4em\relax Cambridge, UK: Camb. Univ. Press, 1997.

\bibitem{Xiu:2002c}
D.~Xiu and G.~E. Karniadakis, ``The {W}iener-{A}skey polynomial chaos for
  stochastic differential equations,'' \emph{{SIAM} Journal on Scientific
  Computing}, vol.~24, no.~2, pp. 619--644, 2002.

\bibitem{Askey:1985}
R.~Askey and J.~Wilson, ``Some basic hypergeometric polynomials that generalize
  jacobi polynomials,'' \emph{Memoirs Amer. Math. Soc.}, vol. 319, pp. 1--55,
  1985.

\bibitem{Smolyak:1963}
S.~A. Smolyak, ``Quadrature and interpolation formulas for tensor products of
  certain classes of functions,'' \emph{Soviet Mathematics Dokl.}, vol.~4, pp.
  240--243, 1963.

\bibitem{Gerstner:1998}
T.~Gerstner and M.~Griebel, ``{N}umerical integration using sparse grids,''
  \emph{Numerical Algorithms}, vol.~18, pp. 209--232, 1998.

\bibitem{Conrad:2013}
P.~Conrad and Y.~Marzouk, ``Adaptive smolyak pseudospectral approximations,''
  \emph{SIAM Journal on Scientific Computing}, vol.~35, no.~6, pp.
  A2643--A2670, 2013.

\bibitem{Patterson:1968}
T.~Patterson, ``The optimum addition of points to quadrature formulae,''
  \emph{Mathematics of Computation}, vol.~22, no. 104, pp. 847--856, 1968.

\bibitem{IEEEdata}
``{P}ower {S}ystems {T}est {C}ase {A}rchive\hfill,''
  \url{http://www.ee.washington.edu/research/pstca/}, accessed: 2014-05-01.

\end{thebibliography}
\end{document}